\documentclass[10pt]{article}
\pdfoutput=1

\usepackage[margin=2.5cm]{geometry}

\usepackage{rotating}
\usepackage{stmaryrd}

\usepackage{cite}

\usepackage{array} 
\usepackage{amssymb}
\usepackage{graphics,graphpap}
\usepackage{graphicx}
\usepackage{color}
\usepackage{graphicx}
\usepackage{dcolumn}
\usepackage{epsfig}
\usepackage{epstopdf}
\usepackage{bbm}
\usepackage{amsmath}
\usepackage{amsfonts}
\usepackage{textcomp}
\usepackage{setspace}
\usepackage{slashed}

\newcommand{\beq}{\begin{eqnarray}}
\newcommand{\eeq}{\end{eqnarray}}

\newcommand{\bmp}{\noindent\begin{minipage}{16cm}}
\newcommand{\emp}{\end{minipage}\vskip 7mm} 

\def\drawbox#1#2{\hrule height#2pt
        \hbox{\vrule width#2pt height#1pt \kern#1pt
              \vrule width#2pt}
              \hrule height#2pt}

\def\Asym#1#2{\vcenter{\vbox{\drawbox{#1}{#2}
              \kern-#2pt 
              \drawbox{#1}{#2}}}}

\def\simge{\mathrel{%
   \rlap{\raise 0.511ex \hbox{$>$}}{\lower 0.511ex \hbox{$\sim$}}}}

\def\simle{\mathrel{
   \rlap{\raise 0.511ex \hbox{$<$}}{\lower 0.511ex \hbox{$\sim$}}}}

\def\s#1{\setbox0=\hbox{$#1$}%
\rlap{\ifdim\wd0>.7em\kern.22\wd0\else\kern.1\wd0\fi /}#1}

\begin{document}

\title{\huge \bf Common origin \\ of reactor and sterile neutrino mixing
}
\maketitle

\begin{center}
Alexander Merle$^a$, Stefano Morisi$^b$, and Walter Winter$^b$\\
\ \\
$^a${\it Physics and Astronomy, University of Southampton, Highfield, Southampton, SO17 1BJ, United Kingdom}\\
$^b${ \it DESY, Platanenallee 6, D-15735 Zeuthen, Germany}
\end{center}

\begin{abstract}
If the hints for light sterile neutrinos from short-baseline anomalies are to be taken seriously, global fits indicate active-sterile mixings of a magnitude comparable to the known reactor mixing. We therefore study the conditions under which the active-sterile and reactor mixings could have the same origin in an underlying flavour model. As a starting point, we use $\mu-\tau$ symmetry in the active neutrino sector, which (for three neutrinos) yields a zero reactor neutrino angle and a maximal atmospheric one. We demonstrate that adding one sterile neutrino can change this setting, so that the active-sterile mixing and non-zero $\theta_{13}$ can be generated simultaneously. From the phenomenological perspective, electron (anti)neutrino disappearance can be easily accommodated, while muon neutrino disappearance can vanish. Even the LSND results can be reconciled if the Majorana phases have very specific values. From the theory perspective, the setting requires the misalignment of some of the flavon vacuum expectation values, which may be achieved in an $A_4$ or $D_4$ flavour symmetry model using extra dimensions. 
\end{abstract}

\section{Introduction}

The current picture of three-flavour neutrino oscillations has been completed by the measurement of a non-zero reactor mixing angle $\theta_{13}$~\cite{An:2012eh}, yielding a self-consistent picture, see Refs.~\cite{GonzalezGarcia:2012sz,Tortola:2012te,Capozzi:2013csa} for global fits. More recently, perhaps even some hint for a CP-violating phase $\delta_{\mathrm{CP}}$ has been already seen in the combination of different experiments~\cite{Capozzi:2013csa}. On the other hand, several anomalies at short baselines indicate that the picture may in fact not be complete, and it thus may have to be extended by one or more sterile neutrinos at the eV-scale (and maybe at other scales, too). In greater detail, evidence for $\bar \nu_\mu \rightarrow \bar \nu_e$ appearance has been found in the LSND experiment~\cite{Aguilar:2001ty}, which has been confirmed by the MiniBooNE experiment in both the neutrino~\cite{AguilarArevalo:2007it} and antineutrino~\cite{Aguilar-Arevalo:2013pmq} modes. These evidences are compatible with one or more extra sterile neutrinos at the eV-scale. On the other hand, recent re-calculations of the reactor $\bar \nu_e$ fluxes~\cite{Mueller:2011nm,Huber:2011wv} are in tension with the corresponding short-baseline disappearance measurements, indicating that a fraction of the electron antineutrinos may have already disappeared into sterile species by oscillations. Finally, somewhat lower event rates than predicted were measured in solar gallium neutrino experiments, yielding a $3\sigma$ indication that electron neutrinos from the Sun are missing, too, which again suggests that these may have partially disappeared into a sterile species~\cite{Giunti:2010zu}. While each of these observations may be interpreted by adding (at least) one extra sterile neutrino, there is a well-known tension between appearance and disappearance data in the global fits, see Refs.~\cite{Kopp:2011qd,Kopp:2013vaa} for recent works.  Several new experiments have been proposed~\cite{Agarwalla:2010zu,deGouvea:2011zz,Rubbia:2013ywa,2013arXiv1304.7127K,Porta:2010zz} to solve these issues and to draw a self-consistent picture, see Ref.~\cite{Abazajian:2012ys} for an extensive review on sterile neutrino phenomenology and experimental prospects.

Due to the increasing amount of experimental indications for eV-scale sterile neutrinos, and also due to slightly heavier (keV-scale) sterile neutrinos being viable candidates for Dark Matter if a suitable production mechanism is used~\cite{Dodelson:1993je,Canetti:2012kh,Shi:1998km,Bezrukov:2009th,Nemevsek:2012cd,King:2012wg,Shaposhnikov:2006xi,Bezrukov:2009yw,Kusenko:2006rh,Petraki:2007gq,Merle:2013wta}, the problem of explaining very light sterile neutrinos has attracted the attention of model builders, see Ref.~\cite{Merle:2013gea} for a recent review.

The basic problem is two-fold:
\begin{enumerate}

\item One has to come up with an explanation for the mass of at least one sterile neutrino being very small (and being protected against radiative corrections), compared to the ``natural'' mass scale for right-handed neutrinos which is thought to be very high (around the scale of grand unification).

\item In addition, one needs to explain the active-sterile mixing $\theta_{i4}$. Depending on the case, this mixing would either need to be sizable, of $\theta_{i4} \sim \mathcal{O}(0.1)$, for eV-sterile neutrinos~\cite{Kopp:2011qd,Kopp:2013vaa} or it should be really tiny, at most of $\theta_{i4} \sim \mathcal{O}(10^{-4})$, for keV-sterile neutrinos~\cite{Watson:2006qb,Abazajian:2001vt,Abazajian:2006jc,Boyarsky:2005us,Dolgov:2000ew,Boyarsky:2006fg,RiemerSorensen:2006fh,Abazajian:2006yn,Boyarsky:2006ag,Boyarsky:2007ge,Loewenstein:2008yi,Watson:2011dw,Loewenstein:2012px}.

\end{enumerate}

Both these requirements are not easy to achieve. Nevertheless, many models have been proposed to solve these problems. A rough classification among the known models distinguishes whether a model attempts to find a unified explanation for both problems, or whether the mechanism to generate a light sterile neutrino mass and the generation of the mixing pattern are separate ingredients. Naturally the former ansatz tends to be much more constrained but, on the other hand, its benefit is being more predictive. Most of the mechanisms to explain light sterile neutrino masses either rely on the principle of suppressing one (or more) sterile neutrino mass eigenvalues or on forcing the natural mass of one sterile neutrino to be zero which is then lifted to a finite but small value by some correction (e.g., by sub-leading terms arising from symmetry breaking).

Models which attempt a simultaneous solution of the light mass problem and of the active-sterile mixing are typically based on flavour symmetries. Known examples include a non-standard $L_e - L_\mu - L_\tau$ lepton number~\cite{Mohapatra:2001ns,Shaposhnikov:2006nn,Lindner:2010wr} or a $Q_6$ symmetry~\cite{Araki:2011zg}, which both force the lightest sterile neutrino to be exactly massless in the symmetry limit but generate a small non-zero mass once the symmetry is broken. Alternatively, a certain mechanism could be used to suppress masses and mixings at the same time, and proposals include the use of the Froggatt-Nielsen mechanism~\cite{Froggatt:1978nt} to explain light sterile neutrinos~\cite{Merle:2011yv} as well as the use of exponential suppressions arising from extra spatial dimensions~\cite{Kusenko:2010ik,Takahashi:2013eva}. Both these proposals have the nice feature that the low energy seesaw mechanism is guaranteed to work, however, they also have the disadvantage that no exact mixing angles can be predicted. Another approach is the use of intermediate scales, which can arise in several extensions of the seesaw mechanism~\cite{Barry:2011wb,Zhang:2011vh,Dev:2012bd}. In general, the most flexible scenarios combine a mass suppression mechanism with a flavour symmetry motivating the mixing, as done for example in the models which use an $A_4$ symmetry in settings where the sterile neutrino mass is suppressed by the Froggatt-Nielsen~\cite{Barry:2011fp,Barry:2011wb}, split seesaw~\cite{Adulpravitchai:2011rq}, or extended seesaw~\cite{Zhang:2011vh} mechanisms. Most of the known models fall into one of the above categories~\cite{Babu:2004mj,Sayre:2005yh,Dias:2005yh,Dinh:2006ia,Cogollo:2009yi,Ma:2009gu,Dias:2010vt,Merle:2012ya,Mavromatos:2012cc,Allison:2012qn,Heeck:2012bz}, although a notable exception exist in which light Dirac-type sterile neutrinos are motivated as composite states~\cite{Grossman:2010iq,Robinson:2012wu}.

In general, it is interesting to ask the question if there are other ways to connect the active and sterile neutrino sectors. We will assume in this paper that some mechanism is at work to explain one very light sterile neutrino -- however, we would like to stress that it is not of great relevance which of the known or yet to be found mechanisms does this job. We then show how the mixings in both active and sterile sectors can be tightly connected in a very simple framework. In particular, the situation considered will allow for predictions in neutrino oscillation experiments which are testable in the very near future.

If the evidence for sterile neutrinos at the eV-scale is to be taken seriously,  global solutions typically point towards
\begin{equation}
U_{e3}\simeq U_{e4} \sim \lambda_C \, ,
\end{equation}
where $U$ is the $4\times 4$ unitary matrix that diagonalises the $4\times 4$ neutrino mass matrix and  $\lambda_C\approx 0.2$. This means that the active-sterile and reactor neutrino mixing angles will be of the same order of magnitude. It is therefore suggestive to investigate scenarios where the active $3 \times 3$ sub-sector of the neutrino mass matrix enforces $\theta_{13}=0$ by a symmetry structure, such as tri-bimaximal (TBM) mixing~\cite{Harrison:2002er} or the $\mu-\tau$ symmetric case~\cite{Fukuyama:1997ky,Mohapatra:1998ka,Ma:2001mr,Lam:2001fb}. For these scenarios well-known flavour symmetry models exist, such as Refs.~\cite{Babu:2002dz,Grimus:2003kq} for the $\mu-\tau$ exchange symmetry case and Refs.~\cite{Ma:2004zv,Altarelli:2005yp,Altarelli:2005yx,Babu:2005se,deMedeirosVarzielas:2006fc} for TBM. By the addition of a sterile species, the mass matrix will be modified and both active-sterile and reactor mixings may be generated. In flavour symmetry models, however, this option turns out to be not that straightforward: the vacuum alignment of the flavon vacuum expectation values (VEVs) prohibits the direct generation of a non-zero $\theta_{13}$, see Refs.~\cite{Barry:2011wb,Barry:2011fp}. We therefore split the problem into two pieces: we first study the requirements for the vacuum alignment in a generic way to produce both active-sterile and reactor mixings of similar magnitudes. Then we discuss the model requirements and how these restrict our generic findings.

We notice that TBM is a special case of the $\mu-\tau$ symmetric case where the solar angle is not  free but \emph{trimaximal}, i.e., $\sin^2\theta_{\rm sol}=1/3$. Since we are interested in studying a possible new origin for the reactor angle independently of the particular value of the solar angle, we consider the general class of $\mu-\tau$ symmetric neutrino mass matrices in this paper. In principle our results could be applied to the subclass of TBM models as well. 

The paper is organised as follow: in Sec.~\ref{sec:method}, we describe our general method and set the stage for the remainder of the paper. Then, in Sec.~\ref{sec:pheno}, we discuss at length our results and their phenomenological consequences. We indicate in Sec.~\ref{sec:theory} how the results can be obtained and sharpened in concrete models, but the discussion of the mathematical details of the models is postponed to Sec.~\ref{sec:models}. We finally conclude in Sec.~\ref{sec:conc}.

\section{\label{sec:method}Method}

Let us consider the $3\times 3$ generic $\mu-\tau$ invariant Majorana neutrino mass matrix given in~\cite{Lam:2001fb},
\begin{equation}\label{eq2}
M_{\mu-\tau} =
\left(
\begin{array}{ccc}
A&B&B\\
B&C &D\\
B&D&C
\end{array}
\right),
\end{equation}
where $A ,B, C,D$ are free parameters. Such a  matrix is diagonalised by the orthogonal matrix
\begin{equation}\label{eq3}
O=
\left(
\begin{array}{ccc}
-c_{12}&s_{12}&0\\
\frac{s_{12}}{\sqrt{2}}& \frac{c_{12}}{\sqrt{2}}&-\frac{1}{\sqrt{2}} \\
\frac{s_{12}}{\sqrt{2}}& \frac{c_{12}}{\sqrt{2}}&\frac{1}{\sqrt{2}} 
\end{array}
\right),
\end{equation}
which has the eigenvector $(0,-1/\sqrt{2},1/\sqrt{2})^T$. This leads to a zero reactor angle and a maximal atmospheric angle, while the solar angle is a function of the parameters  $A, B, C, D$.

We assume only one sterile neutrino  $\nu_s$, and therefore the neutrino mass matrix is given by a $4\times 4$ (symmetric) matrix.\footnote{Introducing several sterile neutrinos does not improve the global fits significantly, at least for a 3+2 instead of a 3+1 model. See e.g. Ref.~\cite{Kopp:2013vaa}.} We furthermore assume the following structure (with the charged leptons being diagonal) in the basis $(\nu_e,\nu_\mu,\nu_\tau,\nu_s)$: 
\begin{equation}\label{mnu44}
 M_\nu^{4\times 4} = 
 \left(
\begin{array}{c|c}
M_{\mu-\tau}&A\\ \hline
A^T&m_s
\end{array}
\right),
\end{equation}
where $m_s$ is the mass contribution of the sterile neutrino assumed to be of the order of $1\,$eV and $A=(a,b,c)^T$ is a $3\times 1$  vector. The vector $A$ can induce mixing effects of the active neutrinos, as discussed in Ref.~\cite{Smirnov:2006bu}. In the limit $A\to 0$ (or if $A$ is an eigenvector of $M_{\mu-\tau}$) the reactor angle is zero, but otherwise the reactor angle can deviate from zero and this deviation is {\it proportional} to the active-sterile mixing, as we will show. In this framework the active-sterile matrix elements $M_{\nu,i4}^{4\times 4}$ (with $i = e, \mu, \tau$) are the origin for the reactor angle. It is noteworthy to add that a model predicting an ``extended $\mu$-$\tau$ symmetry'' could also affect the active-sterile mixings leading to $b=c$. These are not the models we consider in this study, as they would necessarily lead to $\theta_{13}=0$. However, we will demonstrate that $|b|=|c|$ with different phases is compatible with our ansatz, see the discussion in Section~\ref{sec:phases}.

Using the neutrino mass matrix Eq.~\eqref{mnu44}, our purpose is two-fold:
\begin{enumerate}

\item[1)] we want to study if any phenomenological consequences or interplay between the active-sterile mixings emerges from that structure and

\item[2)] we want to investigate the structure of the vector $A$ and which consequences it could have for model builders.

\end{enumerate}

Previous models have studied such interplay in the context of TBM (that is a subclass of our framework~\cite{Barry:2011wb,Barry:2011fp}), but our approach is substantially  different because the reactor angle originates from the sterile sector \emph{only}, while in~\cite{Barry:2011fp} deviations from TBM together plus a sterile neutrino are necessary (in our case, next-to-leading order contributions would \emph{not} be sufficient to generate an acceptable reactor angle).

In this paper, we will embark a numerical analysis of our considerations, supplemented by some analytical approximations. Indeed, it turns out that many aspects are much easier to see numerically than analytically, which simply originates from the fact that, after all, the diagonalisation of a $4\times 4$ mass matrix does involve some complicated formulae. Nevertheless, as we will see, some global tendencies can be seen analytically and, indeed, our general expectations will be confirmed by the numerics.

In our calculation, we first of all assume a general $4\times 4$ neutrino mass matrix by rotating from the mass into the flavour basis assuming Majorana neutrinos,
\begin{equation}
 M_\nu^{4\times 4} = U_{4\times 4}^* {\rm diag}(m_1, m_2, m_3, m_4) U_{4\times 4}^\dagger 
\equiv \begin{pmatrix}
 m_{e 1} & m_{e 2} & m_{e 3} & m_{e 4}\\
 m_{e 2} & m_{\mu 2} & m_{\mu 3} & m_{\mu 4}\\
 m_{e 3} & m_{\mu 3} & m_{\tau 3} & m_{\tau 4}\\
 m_{e 4} & m_{\mu 4} & m_{\tau 4} & m_{s 4}
 \end{pmatrix},
 \label{eq:matrix_expl}
\end{equation}
which is of course symmetric.

In principle there are many different parameterisations of $U_{4\times 4}$ see e.g. \cite{Schechter:1980gr}, since the order of the sub-rotations is arbitrary. Following Refs.~\cite{Donini:2007yf,Donini:2008wz,Adamson:2010wi,Meloni:2010zr,Rodejohann:2011vc}, we choose the parameterisation
\begin{equation}
    \label{equ:3+1param1}
    U_{4\times 4} =
    R_{34}(\theta_{34} ,\, \gamma) \;
    R_{24}(\theta_{24} ,\, \beta) \;
    R_{14}(\theta_{14} ,\, \alpha) \;
    R_{23}(\theta_{23} ,\, \delta_3) \;
    R_{13}(\theta_{13} ,\, \delta_2) \;
    R_{12}(\theta_{12} ,\, \delta_1) \,.
\end{equation}
In Eq.~\eqref{equ:3+1param1}, $R_{ij}(\theta_{ij},\ \varphi)$ are the complex rotation matrices in the $ij$-plane, defined as:
\begin{equation}
[R_{ij}(\theta_{ij},\ \varphi)]_{pq} = \left\{ 
\begin{array}{ll} \cos \theta_{ij} & p=q=i,j\ , \\
1 & p=q \not= i,j\ , \\
\sin \theta_{ij} \ e^{-i\varphi} &	p=i;q=j\ , \\
-\sin \theta_{ij} \ e^{i\varphi} & p=j;q=i\ , \\
0 & \mathrm{otherwise\ .}
\end{array} \right.
\label{eq:rot}
\end{equation}
This means that $\delta_2$ becomes $\delta_{\mathrm{CP}}$ in the three flavour limit. This parameterisation has the advantage that the standard leptonic mixing matrix has to be recovered in the case of vanishing new mixing angles. Note that the order of the 34-24-14-rotations is arbitrary. We chose the 34-angle as the left-most one, which makes it hardest to observe (it affects only $\nu_\tau$-$\nu_s$-mixing). Changing the order here does not change the fact that one of the rotations is difficult to extract. Even though the Majorana phases $(\alpha, \beta, \gamma)$ are absent in the oscillation parameters, they do play an important role for the structure of the mass matrix itself, and in particular for the correlations between different observables. However, there are nevertheless cases in which they have trivial values as predicted by a certain model under consideration. To cover the general tendencies, we will present most of our results first for $(\alpha, \beta, \gamma) = (0, 0, 0)$, in which case even detailed analytical predictions are possible, and we then generalise to arbitrary $(\alpha, \beta, \gamma)$. As will be visible in our plots, the former case will always be a subset of the latter, as to be expected, which confirms the consistency of our numerical procedure.

We therefore have a total of $4 + 6 + 3 = 13$ real parameters, plus potentially 3 Majorana phases:
\begin{itemize}

\item \emph{Four} masses $(m_1, m_2, m_3, m_4)$, where we assume for simplicity normal ordering ($m_1 < m_2 < m_3$) and the fourth (mainly sterile) mass eigenstate to be the heaviest (``$3+1$ scheme'': $m_{1,2,3} < m_4$). A generalisation to other scenarios is straightforward.

\item \emph{Six} mixing angles, three of them ($\theta_{12,13,23}$) describing the ordinary mixing between active neutrinos and three further angles ($\theta_{14,24,34}$) describing the mixing between active and sterile neutrinos.

\item \emph{Three} Dirac phases $\delta_{1,2,3}$, which describe all the CP violation that is potentially measurable in neutrino oscillation experiments.

\item \emph{If applicable: }\emph{Three} Majorana phases $(\alpha, \beta, \gamma)$, which could only be measured in neutrinoless double beta decay~\cite{Rodejohann:2011mu}.\footnote{We leave a detailed study of the predictions for neutrinoless double beta decay for future work.}

\end{itemize}

These parameters can be easily related to short- and long-baseline neutrino oscillation probabilities, see Ref.~\cite{Meloni:2010zr} for details. To leading order in the small mixing angles, the most relevant  short baseline probabilities can be written as:
\begin{align}
\mathcal{P}_{ee} \simeq & 1-\sin^2\left(2 \theta _{14}\right) \, \sin ^2 \Delta_{41} , \label{equ:pee2}\\
\mathcal{P}_{\mu\mu} \simeq & 1- \sin^2\left(2 \theta_{24}\right) \, \sin^2 \Delta_{41}  \label{equ:pmm2} , \\
 \mathcal{P}_{e\mu} = \mathcal{P}_{\mu e} \simeq  & \frac{1}{4} \sin^2 \left( 2 \theta_{14} \right) \, \sin^2 \left( 2 \theta_{24} \right) \,  \sin^2 \Delta_{41} \label{equ:pem2}  , \\
\mathcal{P}_{e\tau}  \simeq  & \frac{1}{4} \sin^2 \left( 2 \theta_{14} \right) \, \sin^2 \left( 2 \theta_{34} \right) \,  \sin^2 \Delta_{41} , \\
\mathcal{P}_{\mu \tau} \simeq &  \frac{1}{4} \sin^2 \left( 2 \theta_{24} \right) \, \sin^2 \left( 2 \theta_{34} \right) \,  \sin^2 \Delta_{41} , \label{equ:pmt2}
\end{align}
where  $\Delta_{ij} \equiv \Delta m_{ij}^2 L/(4 E)$. Note that the CP violating phases and also the light neutrino mass square differences would show up as  corrections to Eqs.~\eqref{equ:pee2} to~\eqref{equ:pmt2} at longer distances. One can easily see in these formulae that, if LSND and MiniBooNE measured the transition in Eq.~\eqref{equ:pem2} which is quadratic in both $\theta_{14}$ and $\theta_{24}$, both electron neutrino disappearance in Eq.~\eqref{equ:pee2} and muon neutrino disappearance in Eq.~\eqref{equ:pmm2} would follow as a consequence. The electron ($\propto \theta_{14}^2$) and muon ($\propto \theta_{24}^2$) neutrino disappearance searches have, so far, not found anything directly, which leads to the well-known tension between appearance and disappearance data. The third mixing angle in our parameterisation, $\theta_{34}$, only enters in $\nu_\tau$ appearance searches, which are much harder to perform because of the high $\tau$ production threshold.

\begin{figure}[tp]
\centering
\begin{tabular}{lr}
\includegraphics[width=7cm]{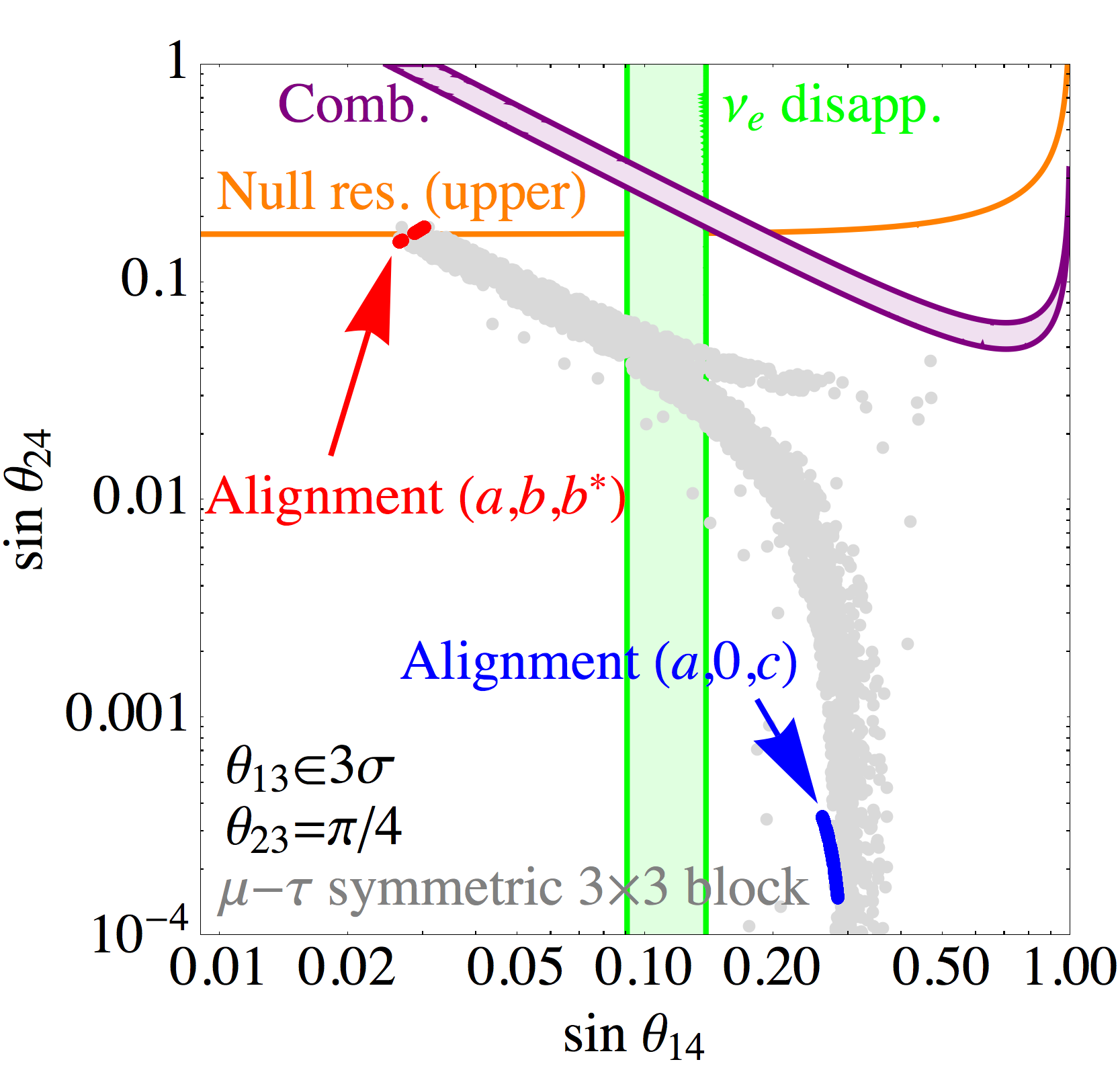} &
\includegraphics[width=7cm]{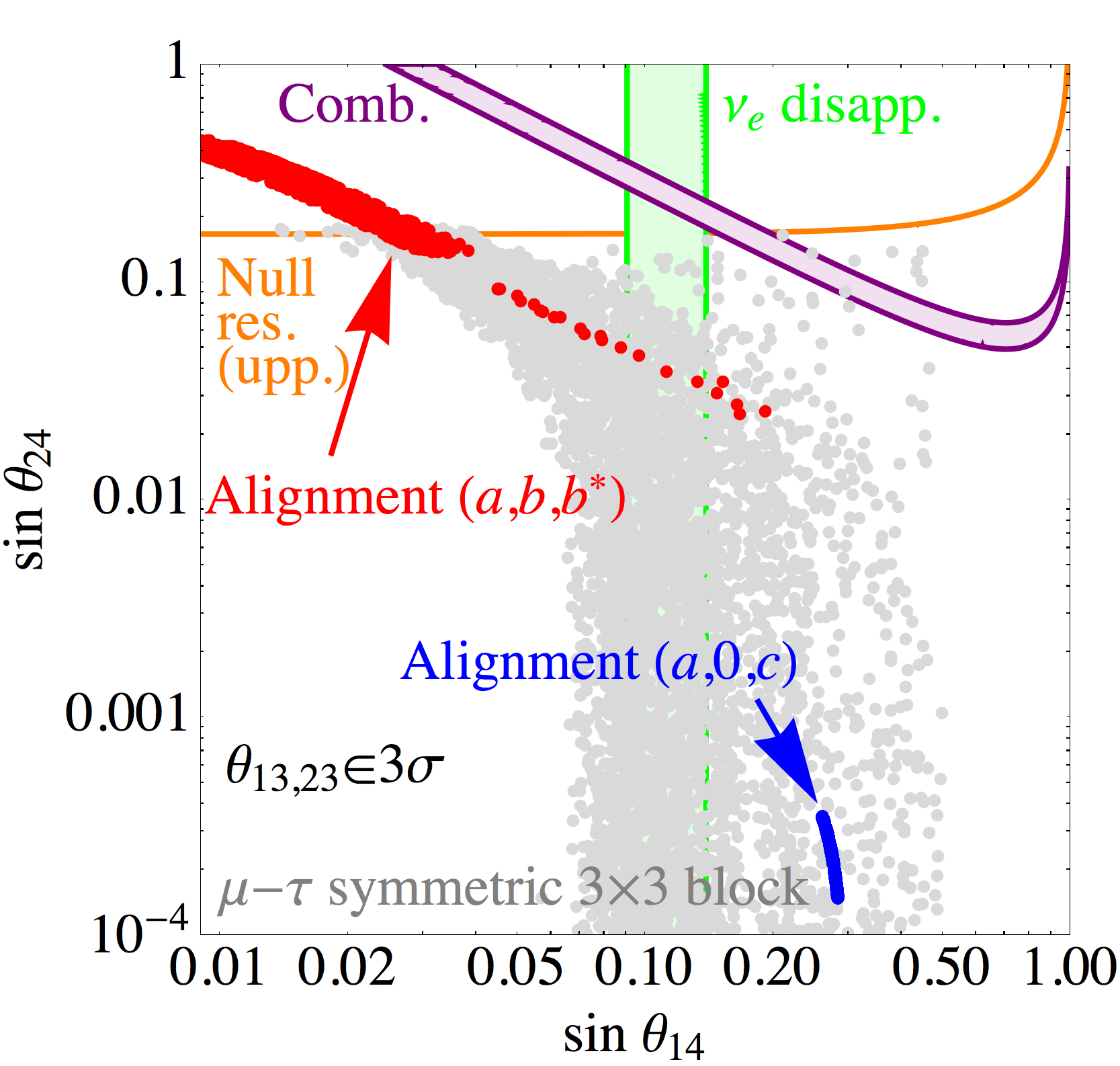}\\
\includegraphics[width=7cm]{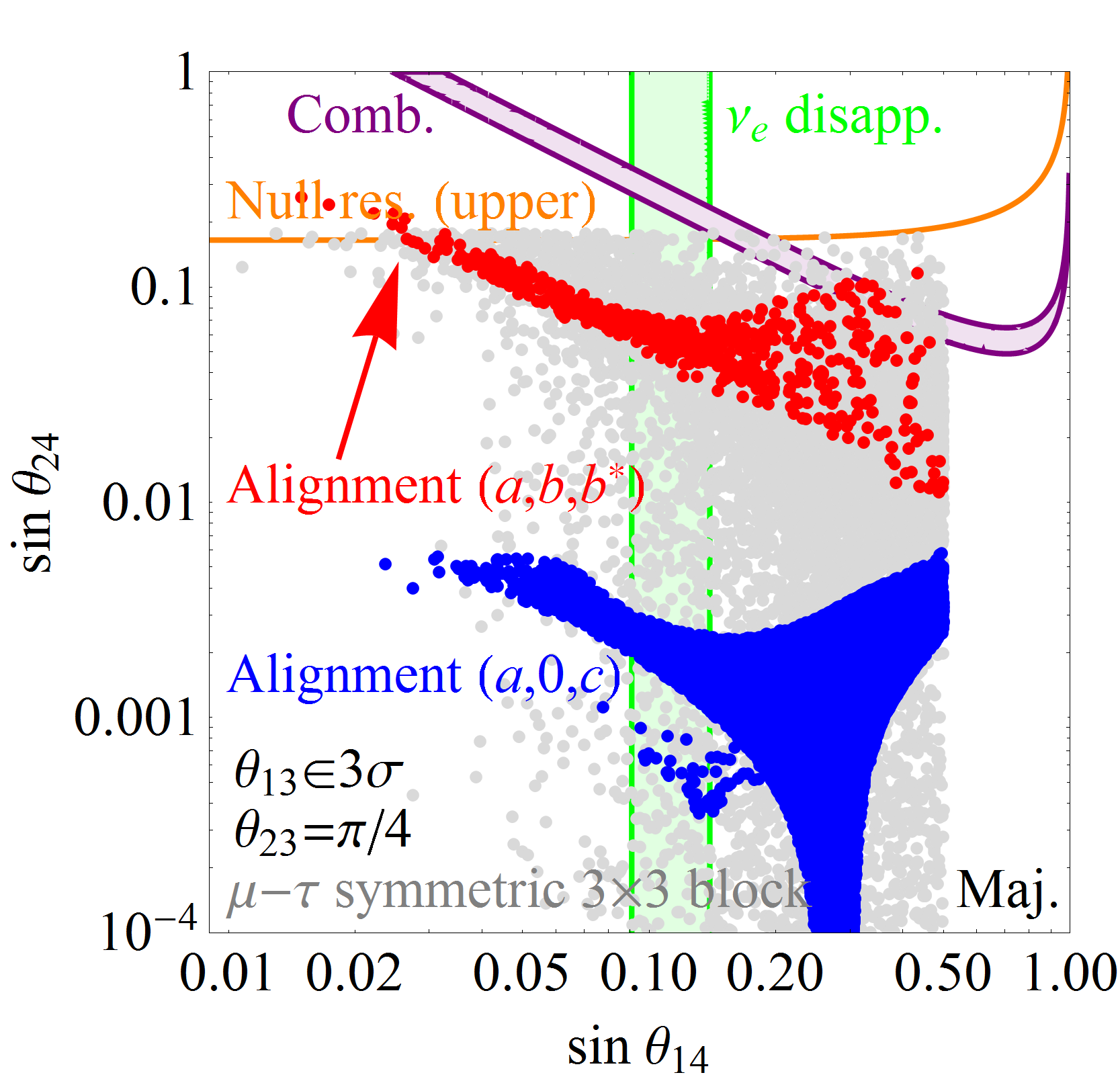} &
\includegraphics[width=7cm]{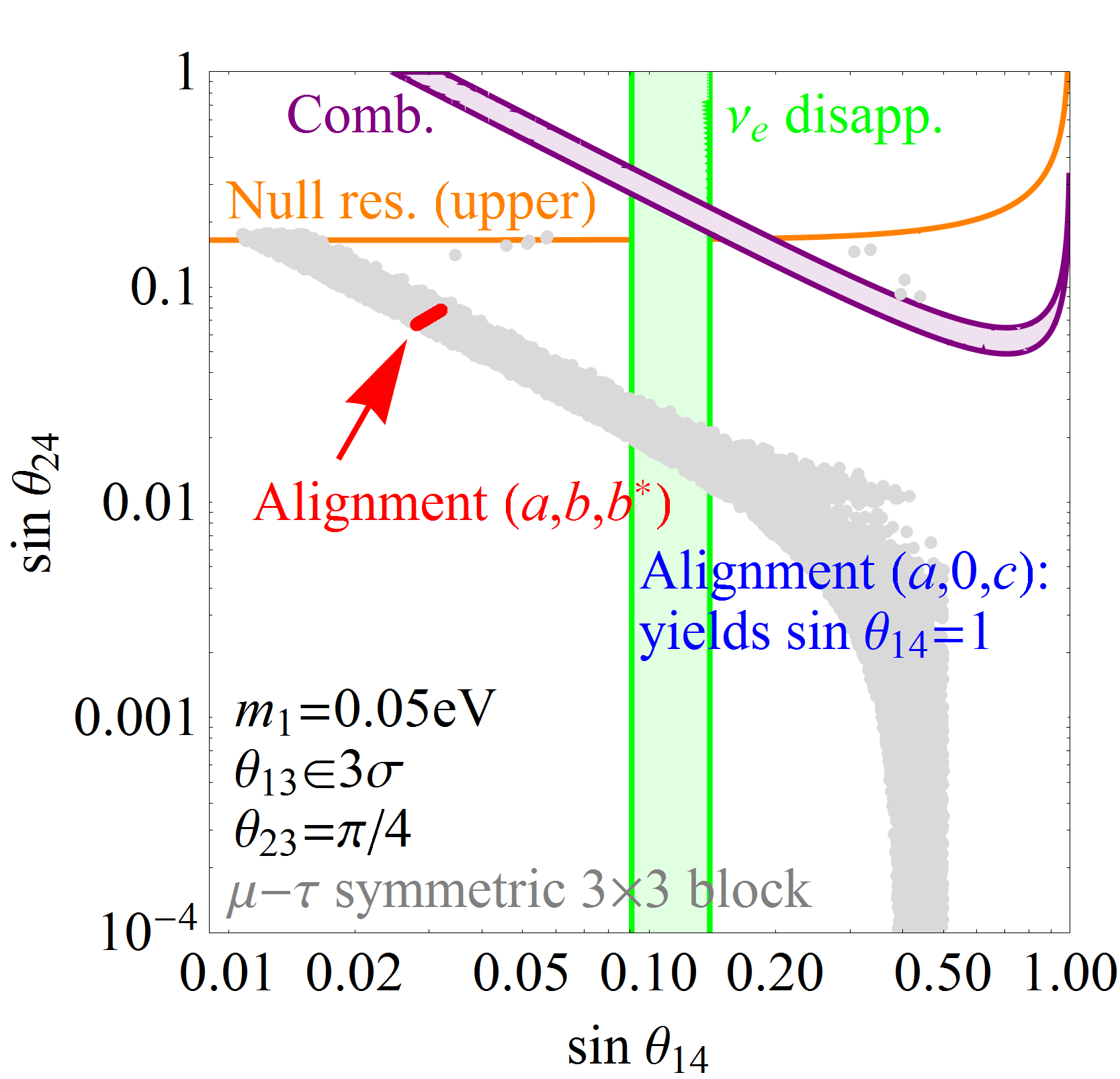}
\end{tabular}
\caption{\label{fig:13-generation1}
Allowed region (gray points) in $\theta_{14}$ and $\theta_{24}$, for $\theta_{13}$ being within its $3\sigma$ range~\cite{GonzalezGarcia:2012sz}. The different panels correspond to zero Majorana phases, $m_1=0$, and $\theta_{23} = \pi/4$ (upper left), zero Majorana phases, $m_1=0$, and $\theta_{23} \in 3\sigma$ (upper right), complex Majorana phases, $m_1=0$,  and  $\theta_{23} = \pi/4$ (lower left), and zero Majorana phases, $m_1=0.05 \, \mathrm{eV}$, and  $\theta_{23} = \pi/4$ (lower right).
 In addition, some example experimental constraints are displayed~\cite{Kopp:2013vaa} (``$\nu_e$ disapp.'' for the region compatible with $\nu_e$ disappearance data, ``Null. res. (upper)'' for the upper limit from $\nu_\mu$ disappearance, ``Comb.'' for the region compatible with combined short-baseline appearance data; see text for details).
It is implied that large $\theta_{14}$ generates large $\theta_{13}$ of the same order, while $\theta_{24}$ could have any value. As visible from the two examples shown, choosing a certain alignment allows to select narrow regions within the general correlated parameter space. For example, one can fix $\theta_{24}$ to be relatively large [$A_4$-like alignment $(a, b, c) = (a, b, b^*)$: red points] or relatively small [$D_4$-like alignment $(a, b, c) = (a, 0, c)$: blue points]. Thus a concrete model can give very clear predictions for the observables.
}
\end{figure}

In our numerical analysis, we have fixed the lightest neutrino mass to be zero, $m_1 = 0$, and the heaviest one to be $m_4 = 1$~eV as an example (with one exception, where we illustrate the effect of $m_1 \neq 0$). Of course we could vary these masses, which would not spoil our principal results but only blur them. We have fixed the other two neutrino masses by imposing the best-fit values~\cite{GonzalezGarcia:2012sz} for the two mass-square differences, $\Delta m_\odot^2$ and $\Delta m_A^2$. Furthermore, we have set $\theta_{12}$ to its best-fit value and we have also set $\theta_{23} = \pi/4$ in order to ensure that the breaking of the $\mu$--$\tau$ symmetry indeed arises from the three parameters $(a, b, c) = (m_{e 4}, m_{\mu 4}, m_{\tau 4})$.\footnote{Note that, alternatively, we could have left $\theta_{23}$ to be a free parameter. Indeed, the contributions from the sterile neutrinos can pull that angle away from its maximal value. However, since the active-sterile mixing angles considered are small after all, it turns out that the resulting interval for $\theta_{23}$ would nevertheless be centered around the maximal value. In particular, the corrections from the sterile sector are \emph{not} large enough to pull this angle to one of its two best-fit values~\cite{GonzalezGarcia:2012sz} in the first or second octant, thus comprising an indirect \emph{signature} of our setting. The correlations shown in our plots would, had we left $\theta_{23}$ free, of course loosen but they would not be wiped out. Thus, for the clarity of the plots and to illustrate the global tendencies of the setting under consideration rather than the influence of experimental uncertainties, we have decided to stick to the choice of $\theta_{23} = \pi/4$. An example of the effect of letting $\theta_{23}$ vary will nevertheless be shown in Fig.~\ref{fig:13-generation1}, upper right panel.} We then generated random values for the parameters $\theta_{13}$ (linear distribution within the $3\sigma$ range of $\sin \theta_{13}$), $\theta_{24}$ (log-scale distribution within $[10^{-5}, 10^{-0.75}]$), and $\delta_2$ (linear distribution within $[0, 2\pi]$). In cases where the Majorana phases $(\alpha, \beta, \gamma)$ have been varied, too, we have also generated random values for each of them, following a linear distribution within $[0, 2\pi]$.

The next step is to impose $\mu$--$\tau$ symmetry onto the upper left $3 \times 3$ block of the full mass matrix $M_\nu^{4\times 4}$ by requiring the two complex equations equations $m_{e 2} = - m_{e 3}$ and $m_{\mu 2} = m_{\tau 3}$ to hold and solving them for the remaining parameters $\theta_{14,34}$ and $\delta_{1,2}$. By this procedure, we have obtained a set of $100,000$ points\footnote{Note that, in the actual plots presented, we show for each region only subsets of the data with a few thousand points each. We have checked that the plots would look practically identical when including all the data so that, had we included all of them, only the file size of the plots would be increased without any significant gain.} which all fulfill the criteria of leading to mass matrices with the desired form of the upper left $3 \times 3$ block and which are phenomenologically valid except for, maybe, their value for $\theta_{13}$, which is exactly what we would like to investigate.

We furthermore have done a similar procedure for two concrete alignments, both of which we will motivate in Sec.~\ref{sec:models} in concrete models. For now, we only observe that for concrete models e.g.\ the family symmetries $A_4$ and $D_4$ can be used. The vector $(a,b,c)$ so far considered can transform as a triplet under $A_4$, or as a singlet plus a doublet in the $D_4$ case. Thus, in concrete models, the vector $(a,b,c)$ is not arbitrary but it is given by the minimisation of the scalar potential invariant under the flavour symmetry of the particular setting considered. Typically, in $A_4$ the following sets of triplet VEV alignments have been studied: 
\begin{equation}
\langle (a,b,c) \rangle  \sim (a,0,0)\,, \quad  a(1,1,1)\,,\quad (a,b,b^*)\,,\quad a(1,4,2)\,,
\label{eq:align_1}
\end{equation}
where the first two alignments are the ones used for TBM, see for instance~\cite{Altarelli:2005yp}, the third alignment is motivated by certain models based on discrete symmetries~\cite{Lavoura:2007dw}, and the fourth one is phenomenologically motivated in~\cite{King:2013iva,King:2013xba}. In the same way in models based on a $D_4$ family symmetry we can have different possibilities for the VEV alignments of the doublet, namely
\begin{equation}
\langle (b,c) \rangle  \sim (b,0)\,, \quad b(1,1)\,,\quad (0,c)\,.
\label{eq:align_2}
\end{equation}


We consider two example alignments: the first one, $(a, b, c) = (a, b, b^*)$, is motivated by an $A_4$ flavour symmetry, while the second one, $(a, b, c) = (a, 0, c)$, can be obtained in models based on $D_4$. From now on, the only important point for us in what concerns phenomenology is that each of these alignments imposes one more complex equation ($c=b^*$ and $b=0$, respectively), which we can use to eliminate two real parameters. Thus, in the generation of numerical mass matrices which fulfill one of the two alignments, we have only generated random values for $\theta_{13}$ and for the Majorana phases if applicable (as in the general case), but numerically solved for $\theta_{24}$ and $\delta_2$.

In the plots, we have also indicated certain bounds and/or experimentally favoured regions for light sterile neutrinos. However, we want to stress that -- at the moment -- not all the data sets stemming from different experiments seem to fit together, see Ref.~\cite{Palazzo:2013me} for a concise discussion. Thus, the best we can do is to show some example bounds and let future experiments decide which of them, if any, are correct. We have therefore extracted three different bounds from Ref.~\cite{Kopp:2013vaa}, where we have in each case used the active-sterile mixing angle regions obtained for $\Delta m_{41}^2 = 1~{\rm eV}^2$.\footnote{For the one case we show where $m_1 = 0.05$~eV instead of zero, this would strictly speaking require a largest mass of $m_4 = 1.00125$~eV, which however is so close to $1$~eV that we have neglected this tiny difference.}

The chosen regions are:

\begin{itemize}

\item \emph{all $\nu_e$ disappearance reactor and solar data} (light green region in our plots; see Fig.~2 in~\cite{Kopp:2013vaa}): $8.24\cdot 10^{-3} \leq |U_{e4}|^2 \leq 1.94\cdot 10^{-2}$, where $U_{e4} = \sin \theta_{14}$,

\item \emph{null results combined from atmospheric and short/long baseline accelerator experiments} (region below the thick orange line in our plots; see Fig.~4a in~\cite{Kopp:2013vaa}): $|U_{\mu 4}|^2 \leq 2.74\cdot 10^{-2}$, where $U_{\mu 4} = \cos \theta_{14} \sin \theta_{24}$,

\item \emph{combined results from $\nu_\mu \to \nu_e$ and $\bar{\nu}_\mu \to \bar{\nu}_e$ appearance data} (light purple region in our plots; see Fig.~7 in~\cite{Kopp:2013vaa}): $2.40\cdot 10^{-3} \leq \sin^2 (2 \theta_{e \mu}) \leq 4.20\cdot 10^{-3}$, where $\sin^2 (2 \theta_{e \mu}) = 4 |U_{e4}|^2 |U_{\mu 4}|^2$.

\end{itemize}

As we had already pointed out, the different data sets available do not seem to fit together at the moment. Accordingly, the setting discussed here cannot be consistent with all of them simultaneously, and thus one should keep in mind that the bounds and favoured regions presented comprise example data sets and are not to be taken fully representative. However, from the current perspective it seems likely that one of them might survive future experimental tests and/or that they will be resolved in terms of the discovery of an unknown systematic error in one type of experiment or maybe even by the discovery of more than one type of light sterile neutrino. On the other hand, no matter which data set is favoured by the reader, our general findings remain correct: it is possible to generate a sizable reactor angle from sterile neutrino contributions to the light neutrino mass matrix.

\section{\label{sec:pheno}Experimental consequences: \\ what phenomenologists are interested in}

We will now discuss our numerical results for the mixing angles. For the moment, let us focus on the general case of a $4\times 4$ light neutrino Majorana mass matrix with a $\mu - \tau$ symmetric upper left $3\times 3$ block, which corresponds to the \emph{light gray} points in all plots. The specific alignments (\emph{red} and \emph{blue} points) will be discussed later on.

\subsection{General case for alignments}

Let us now discuss the correlations which appear -- first for $m_1 = 0$, $(\alpha, \beta, \gamma) = (0,0,0)$, and fixed $\theta_{23}$. In the upper left panel of Fig.~\ref{fig:13-generation1}, we present the correlation between $\sin \theta_{14}$ and $\sin \theta_{24}$, where we have selected the gray points from the lists of numerical mass matrices generated by requiring that $\sin \theta_{13}$ lies within its experimental $3\sigma$ interval~\cite{GonzalezGarcia:2012sz}. The result is a clear correlation between the two active-sterile mixing angles. Indeed, one can see that a sizable (within the $3\sigma$ range) reactor angle $\theta_{13}$ also implies a large mixing angle $\theta_{14}$, i.e., $\sin \theta_{14}\approx 0.02$ to $0.4$, while $\sin \theta_{24}$ can essentially assume all values between $0.2$ and zero. This is a clear tendency we have seen in our data: indeed, had we also included smaller (unphysical) values of $\theta_{13}$ in the plot, we would have seen that $\theta_{14}$ is always of the same order as $\theta_{13}$, while $\theta_{24}$ is in general not strongly constrained. Looking closer, we can see that there exist in fact two branches of the correlation between $\sin \theta_{14}$ and $\sin \theta_{24}$. Notably, the ``upper'' branch also strongly constrains $\sin \theta_{24}$ to be $\gtrsim 0.03$, so that in a large number of cases that angle is also sizable.

Let us now allow for some more freedom, starting with general Majorana phases $(\alpha, \beta, \gamma)$ while we still keep $m_1 = 0$ and $\theta_{23} = \pi/4$. This is the case we will throughout the paper present \emph{below} the corresponding plot with $(\alpha, \beta, \gamma) = (0,0,0)$, so that we should now look at the lower left panel of Fig.~\ref{fig:13-generation1}. As can be seen, the two distinct branches of the correlation are now completely indistinguishable, as can be seen form the gray points.\footnote{Note that, in order not to unnecessarily produce too many unphysical points, we have limited our numerical scan to $|\sin \theta_{i4}| < 0.5$, as can be seen in the plot. This does not present any physical restriction.} However, what remains is nevertheless the tendency of not having a too small $\theta_{14}$, unless $\theta_{24}$ is very large.

For completeness, we have (only for this correlation) also illustrated the effects of relaxing one of the other two assumptions, i.e., either varying $\theta_{23}$ within its $3\sigma$ interval (upper right panel) or taking $m_1 \neq 0$ (lower right panel). For these two cases we have again chosen $(\alpha, \beta, \gamma) = (0,0,0)$, in order not to lose sight of which relaxation has which effect. Starting with the case were $\theta_{23}$ is allowed to be non-maximal, the principal tendencies are not really changed, but the allowed spread of points is increased. This blurs the correlation to some extend (as to be expected). Interestingly, it also leads to at least a few general (light gray) points which are consistent with the region favoured by the combined appearance data (marked by the purple strip, as we will explain below), contrary to the points allowed for $\theta_{23}$ taken to be exactly maximal. Thus, allowing $\theta_{23}$ to vary seems to have, at least at first sight, a similar effect as varying the Majorana phases. A less dramatic effect happens if we instead increase $m_1$ to $0.05$~eV, see the lower right panel. Even though $m_1$ is now considerably different from zero, and in fact $m_1 \sim \sqrt{\Delta m^2_A}$, the qualitative features of the correlation are not destroyed. The two branches are still visible, although not as clearly as for the $m_1 = 0$ case, which comes from the slight change in the shape. The only qualitative change is the few gray points on the upper right of the plot, which did not exist for $m_1 = 0$. More dramatic changes will be present for the alignments, as we will see later.

As already mentioned, we have also displayed the favoured regions from all $\nu_e$ disappearance data (green region in the plots labeled by ``$\nu_e$ disapp.'') and from the combined $e$ to $\mu$ appearance data (purple region in the plots labeled by ``Comb.''), as well as the upper bound from all null results combined (orange thick line in the plots labeled by ``Null res.\ (upper)''). As can be seen, our general region is for fixed Majorana phases incompatible with the combined appearance data if $\theta_{23}$ is maximal (which means in particular that it is incompatible with the LSND results, because the bounds from MiniBooNE are not as stringent for $\Delta m^2 = 1~{\rm eV}^2$). However, the points are easily compatible with all null results (only a very marginal region at the top of the region of interest is cut away by that bound) and also the $\nu_e$ disappearance data can be fitted if $\sin \theta_{14} \sim 0.10$ and $\sin \theta_{24} \sim 0.05$. If $\theta_{23}$ is varied (upper right panel) or if the Majorana are varied (lower left panel), however, there exist at least a few points consistent with the combined appearance region.

\begin{figure}[tp]
\centering
\begin{tabular}{lr}
\includegraphics[width=7cm]{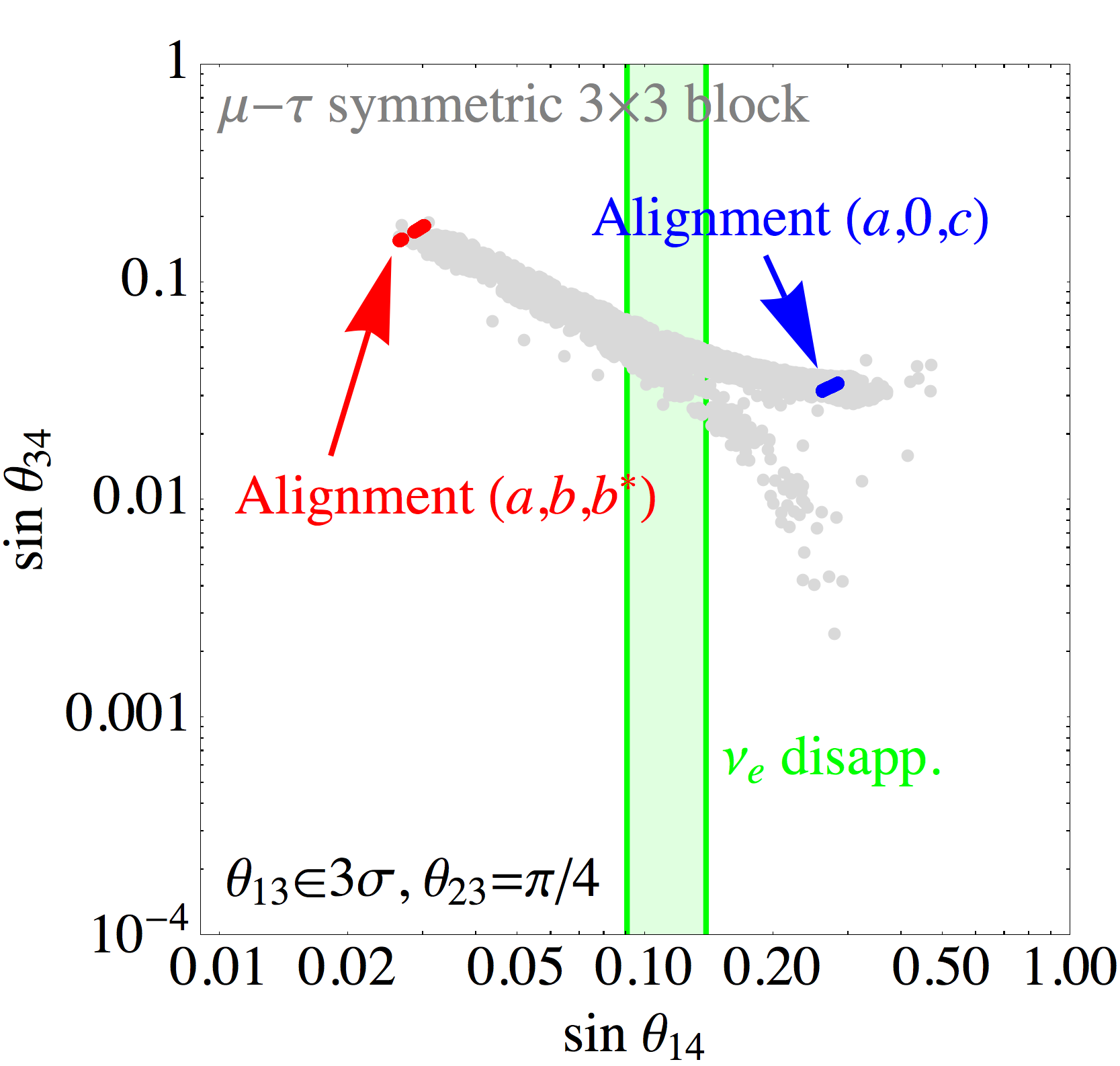} &
\includegraphics[width=7cm]{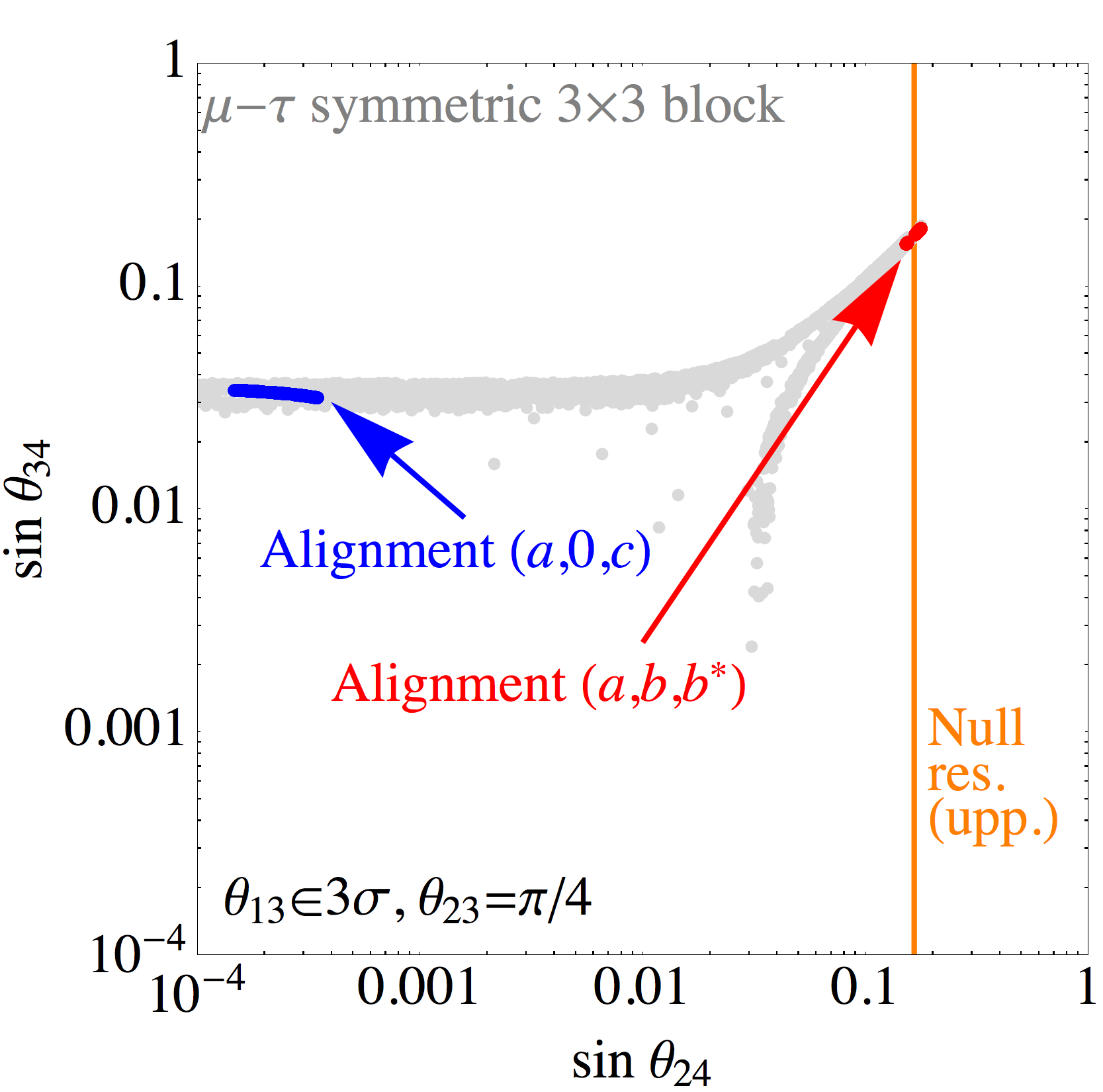}\\
\includegraphics[width=7cm]{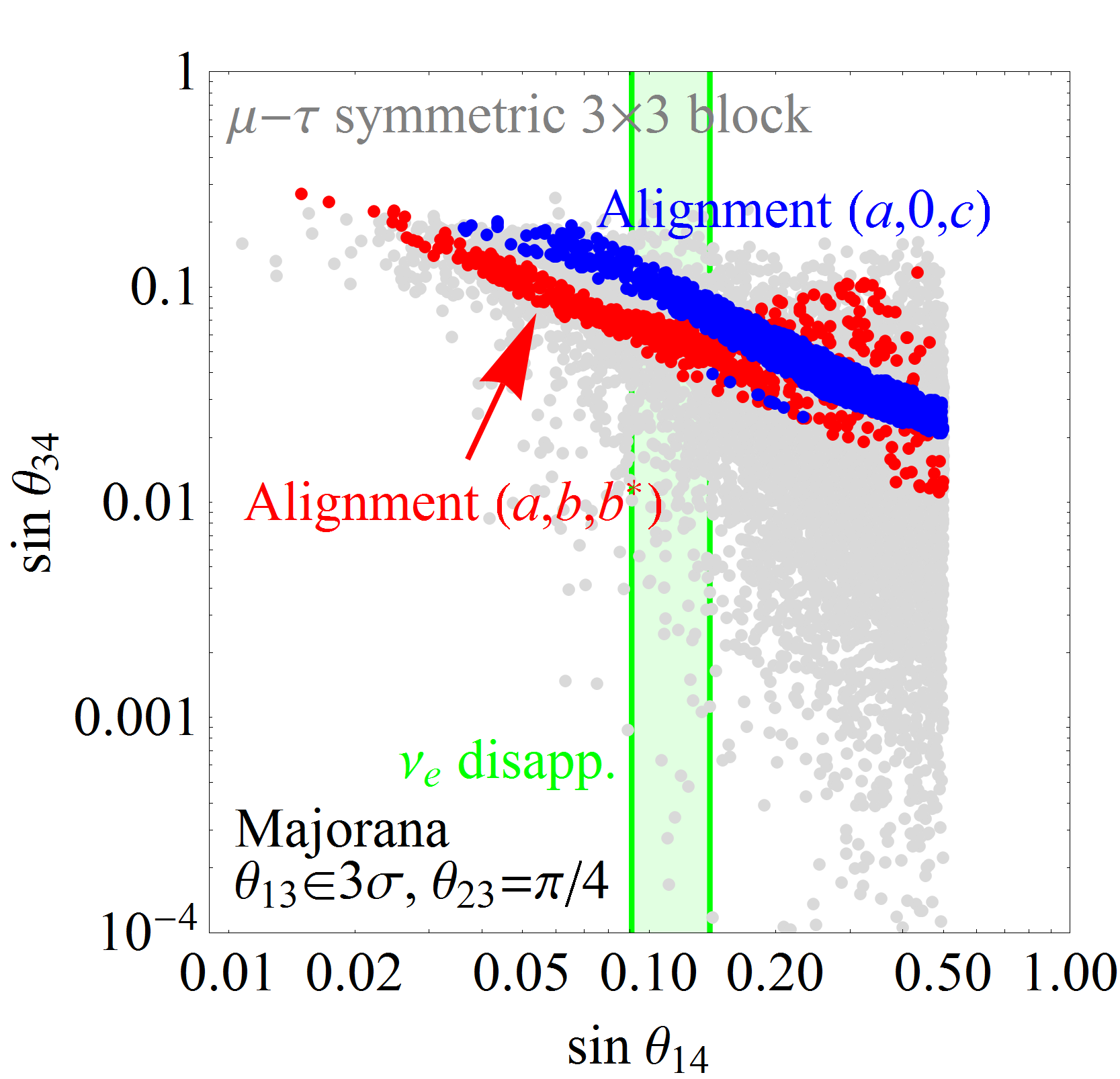} &
\includegraphics[width=7cm]{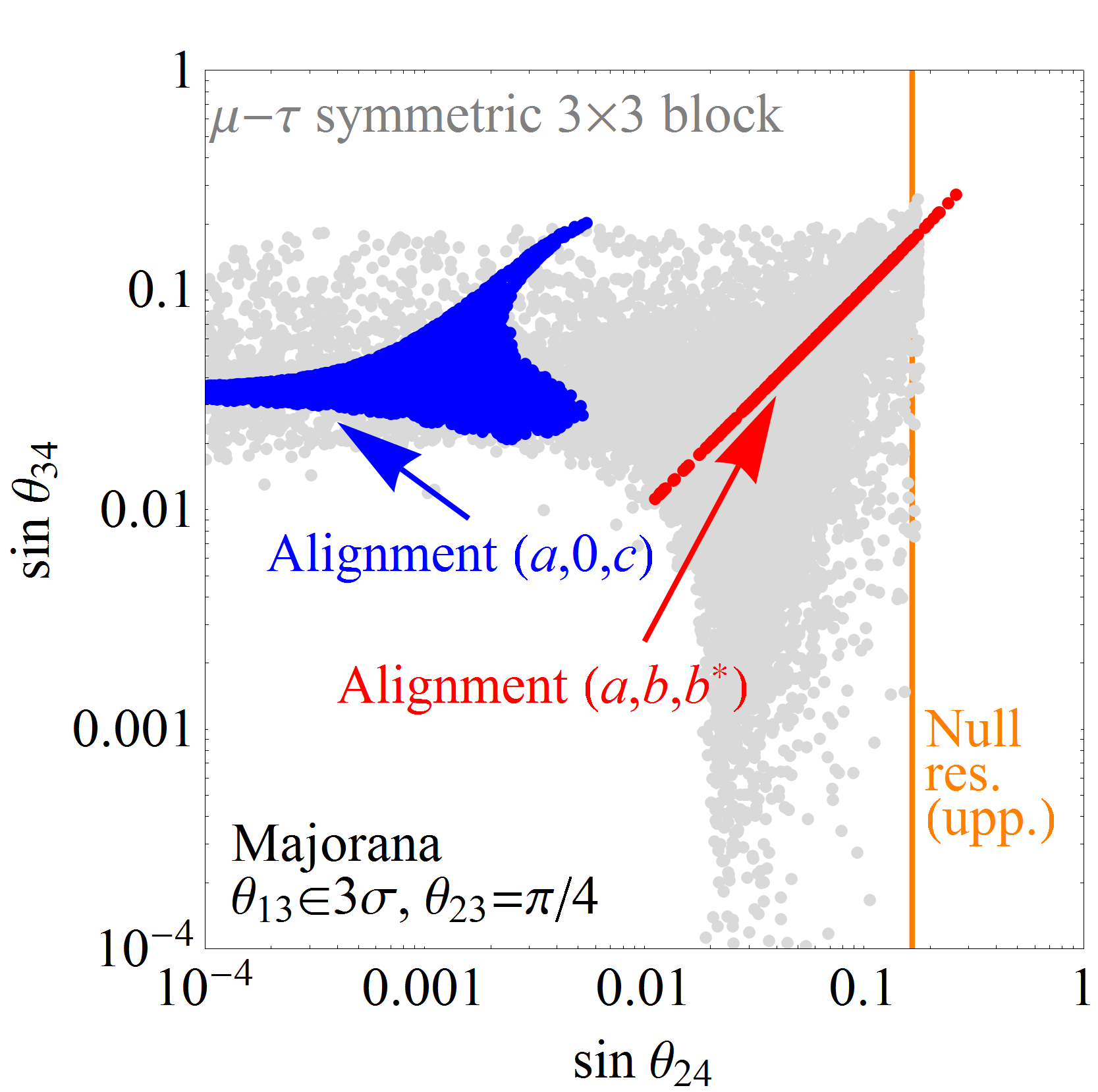}
\end{tabular}
\caption{\label{fig:2434-1314p}
Correlations between $\sin \theta_{14}$ \& $\sin \theta_{34}$ (left panels) and $\sin \theta_{24}$ \& $\sin \theta_{34}$ (right panels). The Majorana phases are chosen to be zero in the upper row and are varied in the lower row. The allowed regions to describe certain data are shown as well, see caption of Fig.~\ref{fig:13-generation1}. As can be seen from both panels, $\theta_{34}$ does have a certain minimal value, while $\theta_{24}$ could be essentially zero (at least in the upper branch of the correlation shown on the right), in consistency with Fig.~\ref{fig:13-generation1}.
}
\end{figure}

Going to Fig.~\ref{fig:2434-1314p}, the correlation between $\sin \theta_{14}$ ($\sin \theta_{24}$) and $\sin \theta_{34}$ is displayed on the left (right) panels. Starting with the left panel and $(\alpha, \beta, \gamma) = (0,0,0)$, it is visible that $\theta_{34}$ is not constrained by the current data, as this angle would correspond to $\nu_\tau \to \nu_s$ transitions which are experimentally hardly accessible. This is why the only favoured region displayed is the green band stemming from the $\nu_e$ disappearance data. The null results do not constrain $\sin \theta_{14}$, and the combined appearance data would not exclude any of the gray points in the $\sin \theta_{14}$--$\sin \theta_{34}$ plane, which is why we have decided not to plot it here. Similarly to the previous case, a clear correlation between $\sin \theta_{14}$ and $\sin \theta_{34}$ is found, again consisting of two distinct branches. However, the difference compared to $\sin \theta_{24}$ is that $\sin \theta_{34}$ (and thus $\theta_{34}$) cannot be arbitrarily small in any branch but is bound to be between roughly $0.2$ and $0.03$ (upper branch) or $0.003$ (lower branch). If the $\nu_e$ disappearance data is to be reproduced, we are forced to have $\sin \theta_{34} \sim 0.05$ (and again $\sin \theta_{14} \sim 0.10$). In the upper right panel, the remaining combination of angles (the correlation between $\sin \theta_{24}$ and $\sin \theta_{34}$) is displayed, which is perfectly consistent with the previous two correlations (one can even make out the correspondences between the different branches). This figure is less favourable in what concerns the experimental bounds, since the upper bound from the null results only appears as a straight line, due to the missing dependence on $\theta_{14}$ in this plot. However, in the region of interest, this does not make a significant difference.

Varying the Majorana phases (lower two panels of Fig.~\ref{fig:2434-1314p}), the correlations are considerably broadened. In particular, it is not possible anymore to distinguish the different branches. Furthermore, also very small values for $\sin \theta_{34}$ are possible in this case. However, it remains true that $\sin \theta_{34}$ and $\sin \theta_{14}$ (or $\sin \theta_{34}$ and $\sin \theta_{24}$) cannot simultaneously be small. This fact can be understood analytically, as we will see later on.

\subsection{Specific alignments}

We will now investigate what changes if we choose a certain vacuum alignment, i.e., a particular form of the vector $A = (a, b, c)$. Such relations are no arbitrary assumptions but they can be derived within concrete models, as we will illustrate later. However, we chose to first present our results to increase the clarity, such that it is easy to see the effect of the alignments, while the inclined reader who is interested in the theoretical details behind the alignments is advised to consult the dedicated Sec.~\ref{sec:models}. 

Looking again at the two leftmost panels of Fig.~\ref{fig:13-generation1}, we have displayed the resulting regions for two different alignments, one of which can be motivated by models based on an $A_4$ symmetry [$(a, b, c) = (a, b, b^*)$, cf.\ Eq.~\eqref{eq:align_1}: red points in the plots] and one of which can be derived from $D_4$ models [$(a, b, c) = (a, 0, c)$, cf.\ Eq.~\eqref{eq:align_2}: blue points in the plots]. The effect of the alignments is immediate: they single out very small patches of the general (light gray) region which, in turn, leads to a high predictivity of the corresponding models. In the case of vanishing Majorana phases, Figs.~\ref{fig:13-generation1} and~\ref{fig:2434-1314p} together tell us that the first alignment (the one with $c = b^*$) predicts $(\sin \theta_{14}, \sin \theta_{24}, \sin \theta_{34}) \sim (0.03, 0.2, 0.2)$ while the second one (where $b=0$) leads to $(\sin \theta_{14}, \sin \theta_{24}, \sin \theta_{34}) \sim (0.3, 2\cdot 10^{-4}, 0.04)$. Indeed, both alignments are highly predictive, so much so that the $A_4$-like (red) alignment (if $\theta_{23} = \pi/4$) is not only incompatible with both the $\nu_e$ disappearance and the combined appearance results (the latter point is not too much of a surprise, given that already the general gray region had been incompatible with this data set), but it is even barely compatible with the not very stringent null results combined. Thus, this alignment case could in fact be excluded very soon. The $D_4$-like (blue) alignment is also only compatible with the null results, but here the predicted value of $\sin \theta_{24}$ is so small that a near-future exclusion of that setting seems more than unlikely.

It is worth to note that varying $\theta_{23}$ does not only spread out the generally allowed set of points, but also the regions allowed for a certain alignment, as can be seen from the upper right panel of Fig.~\ref{fig:13-generation1}. While this effect seems very tiny for the $D_4$-like (blue) alignment, the allowed region for the $A_4$-like (red) alignment is considerably increased. In particular, it is now possible to find red points which match the region allowed by the $\nu_e$ disappearance data, even without varying the Majorana phases. This is very good news, since it means that the red alignment will in fact be a valid possibility if the green region persists, since we cannot expect $\theta_{23}$ to be exactly maximal (and the global fits tell us that a non-maximal value even seems more likely~\cite{GonzalezGarcia:2012sz}, however our setting is unable to reach any of the $\theta_{23}$ best-fit points as both are too far away from $\pi/4$). 

Going back to the case where $\theta_{23}$ is taken to be maximal and comparing the upper left panels of Figs.~\ref{fig:13-generation1} and~\ref{fig:2434-1314p}, it is intriguing that $\sin \theta_{24} \simeq \sin \theta_{34}$ holds for the $A_4$-like (red) alignment. The $D_4$-like (blue) alignment, in turn, leads to a very small angle $\theta_{24}$, whereas $\theta_{34}$ is bound to be on the upper branch of the correlation and thus $\sin \theta_{34} \sim 0.03$. 

What changes if we allow the Majorana phases to vary? As to be expected, also the regions allowed by the alignments are blown up, cf.\ lower left panel of Fig.~\ref{fig:13-generation1} and lower panels of Fig.~\ref{fig:2434-1314p}. The former plot in particular reveals that now, it is not only possible to meet the region favoured by the $\nu_e$ disappearance data for both alignments, but the red alignment can even be consistent with the purple combined $e$ to $\mu$ appearance data. However, the alignments nevertheless clearly reveal certain distinct patterns within the set of gray points. Furthermore, the alignment regions for $(\alpha, \beta, \gamma) = (0,0,0)$ are clearly contained in the more general alignment regions where the Majorana phases are varied, which again confirms the consistency of our numerics. Not too surprisingly, the alignment regions are also blow up for the other correlations, cf.\ lower panels of Fig.~\ref{fig:2434-1314p}. However, what is very remarkable is that the red alignment clearly predicts $\sin \theta_{24} \simeq \sin \theta_{34}$, even if the Majorana phases are varied. This strongly indicates a clear prediction of the red alignment which indeed can be analytically derived as we will see in the next section.

Finally, the most dramatic change of the alignments happens of we choose $m_1 \neq 0$. While the red alignment is only shifted to slightly larger values of $\sin \theta_{14}$, the blue alignment seems to enforce $\sin \theta_{14} \equiv 1$ according to our numerics, and thus violates our condition $\sin \theta_{14} < 0.5$, which is why it does not appear in the plot. This is clearly unphysical, since a maximal active-sterile mixing angle would have been detected already. This is a good example for the predictivity of alignments: while they do allow for some freedom, forcing the mixing angles to be within their physically tolerable ranges might restrict the neutrino masses such that only a certain mass scale between $0$ and $1$~eV is allowed. Turning it round, if the mass scale is known, an alignment can make concrete predictions for at least active-sterile mixings.\\

The principal tendency we wanted to reveal was that \emph{non-trivial sterile mixing can generate a non-zero reactor angle $\theta_{13}$}. This can indeed be seen from the plots in Figs.~\ref{fig:13-generation1} and~\ref{fig:2434-1314p}, which for $(\alpha, \beta, \gamma) = (0,0,0)$ clearly demonstrate that both $\theta_{14}$ and $\theta_{34}$ must be large to generate a sizable reactor angle $\theta_{13}$. On the other hand, $\theta_{24}$ could be small or large, depending on the branch of the correlation. If we allow the Majorana phases to vary, than each of the three active-sterile mixing angles can in principle be small, but not all at the same time: at least one active-sterile mixing angle \emph{must} be large, in order for a sizable reactor angle $\theta_{13}$ to be generated.

\subsection{Analytical understanding}

Let us try to get some analytical understanding of the behaviour shown in the plots. Using Eqs.~\eqref{mnu44}, \eqref{eq:matrix_expl}, \eqref{equ:3+1param1}, and~\eqref{eq:rot}, together with the approximations $m_{1,2} \simeq 0$, $m_3 \simeq \sqrt{\Delta m_A^2}$, an expansion to first order in $s_{14,24,34}$, and neglecting terms like $\sqrt{\Delta m_A^2} s_{13} s_{i4}$ when compared with terms containing $m_4$ and a smaller number of suppressions yields the following approximations for some of the entries in the neutrino mass matrix:
\begin{eqnarray}
 m_{e 2} &\simeq& m_4 e^{-i (\alpha + \beta)} s_{14} s_{24} + \sqrt{\Delta m_A^2} e^{-i(\delta_2 + \delta_3)} s_{13} c_{13} s_{23},\nonumber\\
 m_{e 3} &\simeq& m_4 e^{-i (\alpha + \gamma)} s_{14} s_{34} + \sqrt{\Delta m_A^2} e^{-i \delta_2} s_{13} c_{13} c_{23},\nonumber\\
 m_{\mu 2} &\simeq& \sqrt{\Delta m_A^2} e^{-2 i \delta_3} c_{13}^2 s_{23}^2,\nonumber\\
 m_{\tau 3} &\simeq& \sqrt{\Delta m_A^2} c_{13}^2 c_{23}^2.
 \label{eq:12132233}
\end{eqnarray}
As already mentioned in Sec.~\ref{sec:method}, the conditions for a $\mu - \tau$ symmetric upper left $3\times 3$ block are:
\begin{equation}
 m_{e 2} = - m_{e 3}\ \ \ {\rm and}\ \ \ m_{\mu 2} = m_{\tau 3}.
 \label{eq:mu-tau}
\end{equation}
Applying the latter condition to Eq.~\eqref{eq:12132233}, one obtains $e^{-2 i \delta_3} s_{23}^2 \simeq c_{23}^2$, which immediately implies $\delta_3 \simeq 0$ and
\begin{equation}
 \sin \theta_{23} \simeq \cos \theta_{23} \simeq \frac{1}{\sqrt{2}}\ \ \ \Rightarrow\ \ \ \theta_{23} \simeq \frac{\pi}{4}.
 \label{eq:23max}
\end{equation}
This confirms that $\theta_{23}$ should be very close to maximal, as we had already mentioned. Then, using the first condition from Eq.~\eqref{eq:mu-tau} and inserting $\delta_3 \simeq 0$ and $s_{23} \simeq c_{23} \simeq 1/\sqrt{2}$, one obtains
\begin{equation}
 s_{14} (s_{24} e^{-i \beta} + s_{34} e^{-i \gamma}) \simeq - \frac{\sqrt{2 \Delta m_A^2}}{m_4} e^{-i (\alpha - \delta_2)} s_{13} c_{13} \approx 0.01,
 \label{eq:branches}
\end{equation}
where we have in the final step inserted the best-fit values of the remaining oscillation parameters as well as $m_4 = 1$~eV and $\delta_2 \simeq \pi$, the latter being implied for vanishing Majorana phases, $(\alpha, \beta, \gamma) = (0,0,0)$. It is this equation which teaches us quite a bit about the plots presented in Figs.~\ref{fig:13-generation1} and~\ref{fig:2434-1314p}. First of all, as we had anticipated in Sec.~\ref{sec:method}, the equation proves our central point: up to terms of $\mathcal{O}(s_{13}^3)$ arising from the cosine of $\theta_{13}$, it is indeed true that \emph{the reactor mixing is proportional to the active-sterile mixing}.

Let us again start with the case of vanishing Majorana phases, $(\alpha, \beta, \gamma) = (0,0,0)$. Then, in particular one would necessarily switch off the reactor mixing angle $\theta_{13}$ if either $\theta_{14}$ or $\theta_{24,34}$ were zero. Second, in order for the numerical version of Eq.~\eqref{eq:branches} to hold, $s_{14}$ must be of $\mathcal{O}(0.01)$ or even larger, which is consistent with the limit $\sin \theta_{14}\gtrsim 0.02$ obtained from the plots. Third, only one of $s_{24,34}$ can be small. This fact explains the two branches in Figs.~\ref{fig:13-generation1} and~\ref{fig:2434-1314p}: in the upper left panel of Fig.~\ref{fig:13-generation1} (Fig.~\ref{fig:2434-1314p}), the upper branch is obtained for sizable $s_{24}$ ($s_{34}$), while the lower branch allows for very small values of $s_{24}$ (significantly smaller values of $s_{34}$). The overlap regions of each pair of branches indicate the both angles $s_{24,34}$ are sizable. The differences between $s_{24}$ and $s_{34}$ can be attributed to the sub-leading terms neglected in Eq.~\eqref{eq:12132233}. These considerations basically remain true for general phases $(\alpha, \beta, \gamma)$. The absolute value of the right-hand side of Eq.~\eqref{eq:branches} will be sizable for non-zero reactor angle $\theta_{13}$ and, while the terms in parentheses on the right-hand side could in principle cancel even for large $\theta_{24} = \theta_{34}$ if $\beta = \gamma + \pi$, they cannot sum up to a large number if all angles are small. Thus, even in the general case, a relatively large $\theta_{13}$ enforces a large $\theta_{14}$ and either $\theta_{24}$ or $\theta_{34}$ to be sizable, too.

We can also get some analytical understanding of the effect of the alignments: again using Eqs.~\eqref{mnu44}, \eqref{eq:matrix_expl}, \eqref{equ:3+1param1}, and~\eqref{eq:rot}, it is easy to see that in the limit $m_{1,2,3} \ll m_4$, one obtains
\begin{eqnarray}
 a &\simeq& m_4 e^{-i \alpha} \sin \theta_{14}\cdot \cos \theta_{14} \cos \theta_{24} \cos \theta_{34},\nonumber\\
 b &\simeq& m_4 e^{-i \beta} \sin \theta_{24}\cdot \cos^2 \theta_{14} \cos \theta_{24} \cos \theta_{34},\nonumber\\
 c &\simeq& m_4 e^{-i \gamma} \sin \theta_{34}\cdot \cos^2 \theta_{14} \cos^2 \theta_{24} \cos \theta_{34}.
 \label{eq:bc}
\end{eqnarray}
The $D_4$-like (blue) alignment requires $b=0$ and we thus know that $\cos^2 \theta_{14} \sin (2 \theta_{24}) \cos \theta_{34}\simeq 0$. Furthermore, $\cos \theta_{14}$ and $\cos \theta_{34}$ cannot be zero since $\theta_{14, 34}$ must be somewhat small. This immediately leads to $\sin (2 \theta_{24}) \simeq 0$ and thus requires a very small angle $\theta_{24}$, which is perfectly consistent with our numerical results, cf.\ Fig.~\ref{fig:13-generation1} and right panel of Fig.~\ref{fig:2434-1314p}, even in the general case of arbitrary Majorana phases. Using a similar approximation as in Eq.~\eqref{eq:12132233}, we could alternatively have derived
\begin{equation}
 b \simeq m_4 e^{-i \beta} s_{24} - \sqrt{\frac{\Delta m_A^2}{2}} \left[e^{-i (\alpha  - \delta_2)} s_{13} c_{13} s_{14} + c_{13}^2 \frac{s_{24} e^{i\beta} + s_{34} e^{i\gamma}}{\sqrt{2}} \right] \stackrel{!}{=} 0,
 \label{eq:b-alternative}
\end{equation}
where we have already inserted $s_{23} \simeq c_{23} \simeq 1/\sqrt{2}$. For vanishing $(\alpha, \beta, \gamma)$, which also implies $\delta_2 \simeq \pi$, this equation cannot be solved for $s_{34} \simeq 0$, since the left-hand side would then necessarily be positive. However, in the ``opposite'' limit, $s_{24} \simeq 0$, one can easily find a solution $s_{34} \approx \sqrt{2} \tan \theta_{13} s_{14}$ which, inserting the best-fit value for $\theta_{13}$, implies that $s_{14} \approx 5 s_{34}$. Looking at the upper left panel of Fig.~\ref{fig:2434-1314p}, this relation indeed seems to be approximately fulfilled for the blue alignment. Glancing at the figures with arbitrary Majorana phases, it is visible that the general tendency of avoiding $s_{34} \simeq 0$ again remains true for the blue alignment, although the allowed regions of course open up a little.

For the $A_4$-like (red) alignment, in turn, $c = b^*$ is enforced, where
\begin{equation}
 c \simeq m_4 e^{-i \gamma} s_{34} - \sqrt{\frac{\Delta m_A^2}{2}} \left[e^{-i (\alpha  - \delta_2)} s_{13} c_{13} s_{14} + c_{13}^2 \frac{s_{24} e^{i\beta} + s_{34} e^{i\gamma}}{\sqrt{2}} \right] \stackrel{!}{=} 0,
 \label{eq:c-alternative}
\end{equation}
which immediately implies that $b - m_4 e^{-i \beta} s_{24} \simeq c - m_4 e^{-i \gamma} s_{34}$. For $(\alpha, \beta, \gamma) = (0,0,0)$, combining Eqs.~\eqref{eq:b-alternative} and~\eqref{eq:c-alternative} results in $\sin \theta_{34} \simeq \tan \theta_{24}$, which is approximately equal to $\sin \theta_{24}$ due to the angle $\theta_{24}$ being small. Thus, this alignment leads to $\sin \theta_{34} \simeq \sin \theta_{24}$ and it is exactly that part of the general region which is numerically predicted by the red alignment, cf.\ upper right panel of Fig.~\ref{fig:2434-1314p}. Furthermore, when using Eq.~\eqref{eq:branches} in addition, one can also see that $s_{14} s_{24,34} \sim 0.005$. In in the upper left panel of Fig.~\ref{fig:13-generation1} (Fig.~\ref{fig:2434-1314p}), one can read off $s_{14} \sim 0.03$ and $s_{24} \sim 0.2$ ($s_{34} \sim 0.2$) for the red alignment, which is in good agreement with our analytical estimate. Remarkably, the prediction $\sin \theta_{34} \simeq \sin \theta_{24}$ for the red alignment remains perfectly valid even in the case of non-vanishing $(\alpha, \beta, \gamma)$, cf.\ lower right panel of Fig.~\ref{fig:2434-1314p}. This can be seen most easily by approximating $\sqrt{\Delta m_A^2} \approx 0$ in Eqs.~\eqref{eq:b-alternative} and~\eqref{eq:c-alternative}, which is justified because this quantity is always multiplied by the sines of angles which are not too large. Then, $c = b^*$ and thus $|b| = |c|$ immediately implies $\sin \theta_{34} \simeq \sin \theta_{24}$, which confirms our numerical results.

\section{\label{sec:theory}Results for the mass matrix:\\ what model builders want to know}

The next question to ask is about the concrete connection between the mass matrix entries $a = m_{e4}$, $b = m_{\mu 4}$, and $c = m_{\tau 4}$ and the active-sterile mixing angles. These are the results which are interesting for model builders, because they will reveal which alignments, i.e., ``directions'' of the complex vector $(a,b,c)$ are compatible with the allowed regions in our plots. Again, we will first of all present our general results, i.e., the elements $(a,b,c)$ are arbitrary as long as the resulting points are experimentally valid, and afterwards we will discuss more specifically how certain alignments, i.e., special choices of $(a,b,c)$ as derived within the framework of flavour models which can dramatically sharpen the predictions.\\

\begin{figure}[tp]
\centering
\begin{tabular}{lr}
\includegraphics[width=7cm]{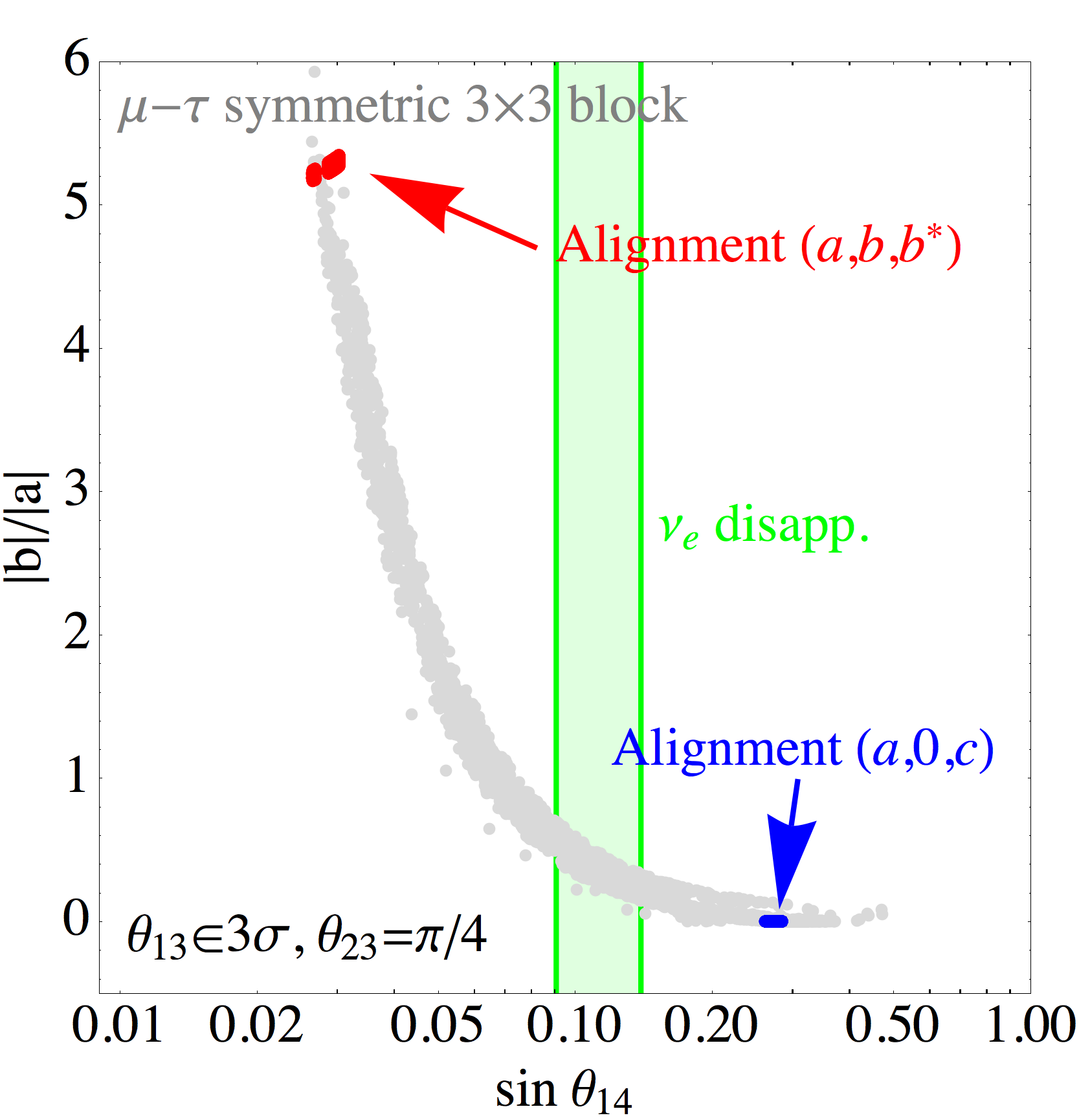} &
\includegraphics[width=7cm]{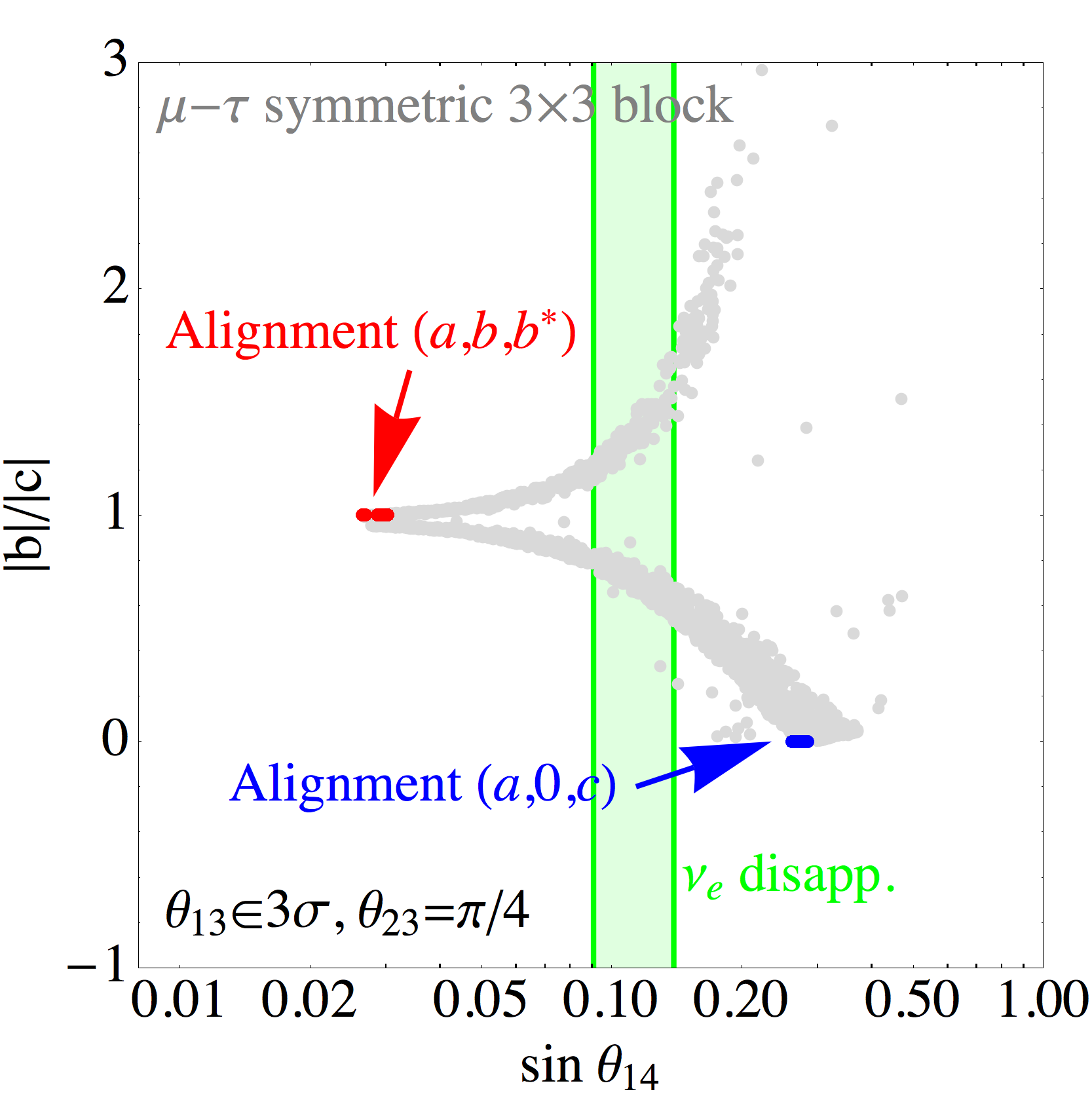}\\
\includegraphics[width=7cm]{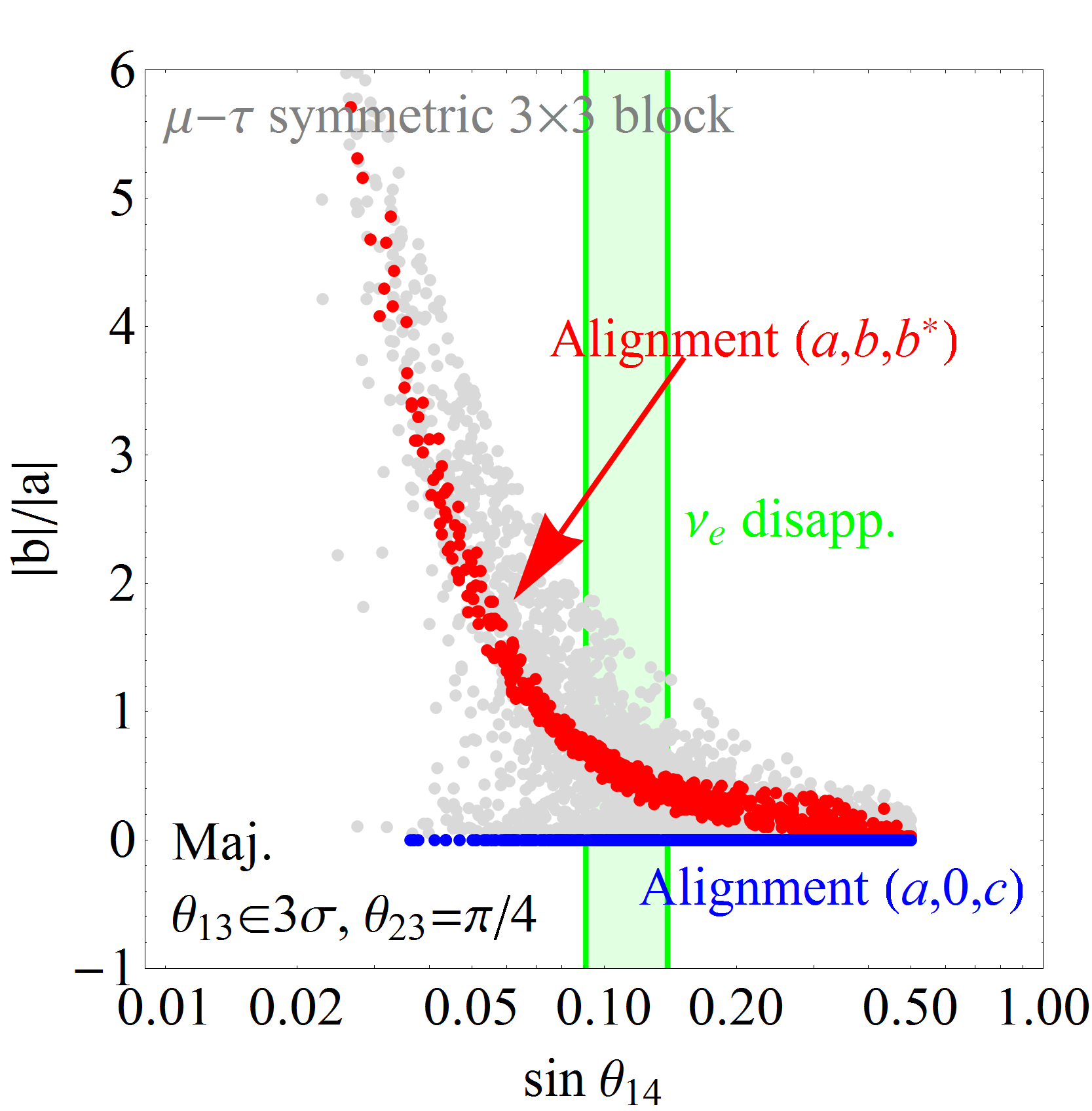} &
\includegraphics[width=7cm]{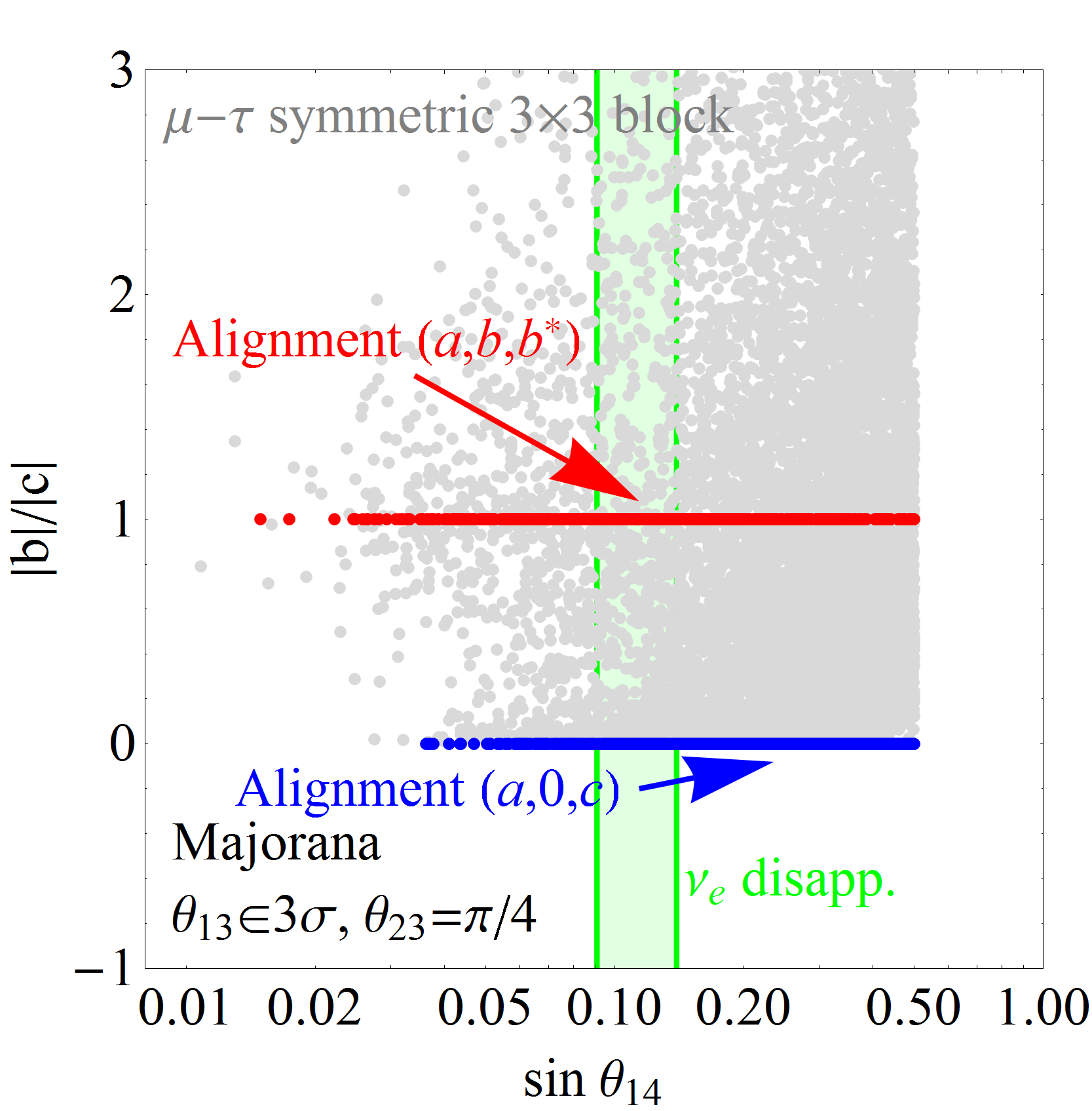}
\end{tabular}
\caption{\label{fig:elements_14}
Correlations between $\sin \theta_{14}$ and the ratios $|b|/|a|$ and $|b|/|c|$ in the left and right columns, respectively. The Majorana phases are chosen to be zero in the upper row, and are varied in the lower row. Again, the connection between the different quantities is very clear for vanishing Majorana phases: a certain value of $|b|/|a|$ corresponds to a very definite value of $\theta_{14}$ and the same is true for $|b|/|c|$, although there exist two different possibilities for that quantity for a given $\theta_{14}$. When varying the phases, some correlation persists for $|b|/|a|$, while it is wiped out completely for $|b|/|c|$.
}
\end{figure}

\subsection{Correlations between observables and absolute values of the alignments}

As examples we display in Fig.~\ref{fig:elements_14} the correlations between $\sin \theta_{14}$ and $|b|/|a|$ (left panels) and $|b|/|c|$ (right panels).\footnote{Note that we could have chosen $\sin \theta_{24,34}$ instead, but these plots would not add anything significant.} Starting with the correlation of $|b|/|a|$ and again assuming vanishing $(\alpha, \beta, \gamma)$ for the start, cf.\ upper left panel, it is clearly visible that there is practically a one-to-one correspondence between the value of $|b|/|a|$ and that of $\sin \theta_{14}$. Naturally, $|b|/|a|$ is bound to be positive, but it can be nearly zero for large values of $\sin \theta_{14} \gtrsim 0.2$. Lower values of $\sin \theta_{14}$ quickly increase the ratio $|b|/|a|$ to roughly $5$ for the smallest possible value of $\sin \theta_{14} \sim 0.03$. This can again be understood analytically: using Eq.~\eqref{eq:bc} for $\alpha = \beta = 0$, it is easy to see that $|b|/|a| \simeq \sin \theta_{24}/\tan \theta_{14}$. This is clearly reflected in the curve depicted in the upper left panel of Fig.~\ref{fig:elements_14}. Imposing the restriction from the example data sets (which is only the $\nu_e$ disappearance data in this case), one can see that $|b| \sim 0.1 |a|$ or, more generally, $|b| \ll |a|$ is enforced. Allowing for varying Majorana phases (cf.\ lower left panel), the correlation between $|b|/|a|$ and $\sin \theta_{14}$ gets broader, but it is not wiped out. In particular for large values of $|b|$, there is still a rough one-to-one correspondence left. However, for very small values of $|b|$, such as enforced by the blue alignment, the allowed range for $\sin \theta_{14}$ becomes quite large. 

Looking at the alignments for $(\alpha, \beta, \gamma) = (0,0,0)$, the $A_4$-like (red) alignment corresponds to the upper range of $|b|/|a| \sim 5$, due to $\sin \theta_{14}$ being very close to its lowest predicted value in that case. Note that, in this alignment, a relation like $|b| \gg |a|$ has never been imposed, but it is instead a consequence of the tightness of the parameter space and thus a reflection of the predictivity of the concrete alignment. The $D_4$-like (blue) alignment in turn enforces $b=0$ (and hence trivially $|b|/|a|=0$), which is confirmed by the resulting points and thus comprises a sanity check of our numerical calculations. As before, $\sin \theta_{14}$ is slightly smaller than 0.3 for this alignment. Varying the Majorana phases allows the red alignment to go much further down to lower values of $|b|/|a|$, however, a clear one-to-one correspondence between $|b|/|a|$ and $\sin \theta_{14}$ remains present to some extend. As anticipated for the blue alignment, having $|b| = 0$ opens up many possibilities for $\sin \theta_{14}$, which can now be as small as about $0.04$.

On the right panels, the correlation between $\theta_{14}$ and $|b|/|c|$ is displayed. For vanishing $(\alpha, \beta, \gamma)$, cf.\ upper right panel, it consists of two branches. For very small values of $\theta_{14}$ (close to the lowest value possible, $\sin \theta_{14}\sim 0.03$), both branches meet and enforce $|b| \simeq |c|$. For larger values of $\theta_{14}$, however, the two branches split and enforce $|b|\neq |c|$. For the upper branch, a rough bound of $|c| < |b| \lesssim 2.5 |c|$ is visible, although there are a few points above that boundary. For the lower branch, in turn, there is no limit except for the trivial one, $|b|/|c|\geq 0$. Note that this curve cannot be understood as easily on analytical grounds: Eq.~\eqref{eq:bc} only implies that $|b|/|c| \simeq \tan \theta_{24}/\sin \theta_{34}$, and thus the dependence on $\theta_{14}$ must arise from sub-leading terms. Imposing the $\nu_e$ disappearance data enforces either $|b| \sim 1.25 |c|$ or $|b| \sim 0.8 |c|$. While these tendencies are nicely visible, the lower right panel of Fig.~\ref{fig:elements_14} reveals that the correlation between $\theta_{14}$ and $|b|/|c|$ is practically wiped out for general phases $(\alpha, \beta, \gamma)$. Thus, in this case, getting useful information is only possible for models which predict fixed values of the Majorana phases. 

The picture looks similar for the alignments. As already mentioned, in the case of vanishing Majorana phases the $A_4$-like (red) alignment yields quite a small $\theta_{14}$. Here we can see why: $(a, b, c) = (a, b, b^*)$ trivially imposes $|b| = |c|$, and this is only possible for small $\theta_{14}$, as can be seen from the gray points. The $D_4$-like (blue) alignment in turn leads to a pretty large $\theta_{14}$. Also this is clear from this figure: $(a, b, c) = (a, 0, c)$ requires $|b|/|c|=0$, which can only be fulfilled if $\theta_{14}$ is large enough. Both these tendencies get practically wiped out if the phases are allowed to have arbitrary values, in which case the alignments do not give more of a prediction than the trivial ones, i.e., $|b| / |c| = 1$ (red) and $|b| / |c| = 0$ (blue).

\subsection{Correlations between observables and phases of the alignments}
\label{sec:phases}

\begin{figure}[tp]
\centering
\begin{tabular}{lr}
\includegraphics[width=7cm]{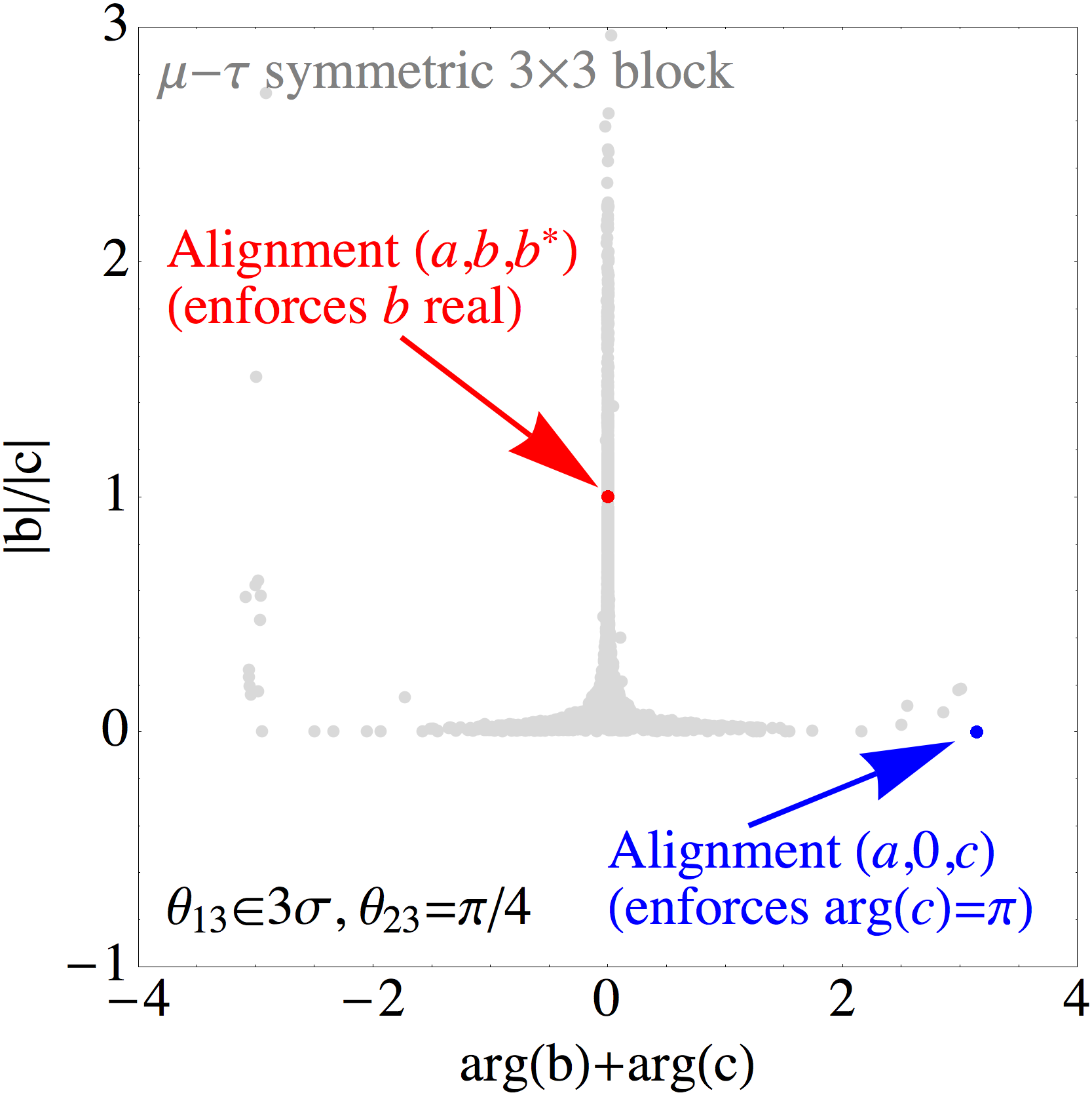} & \includegraphics[width=7cm]{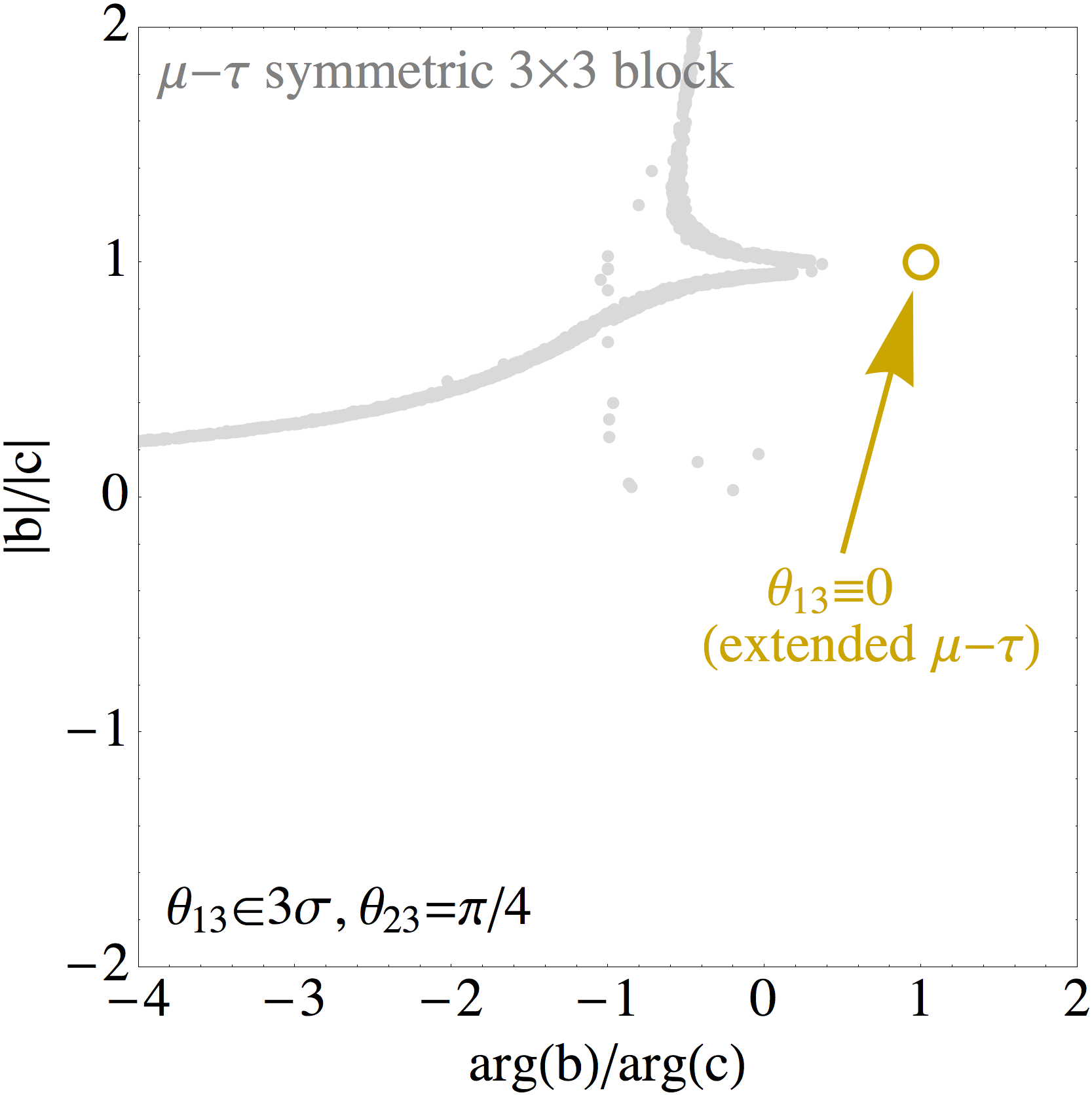}\\
\includegraphics[width=7cm]{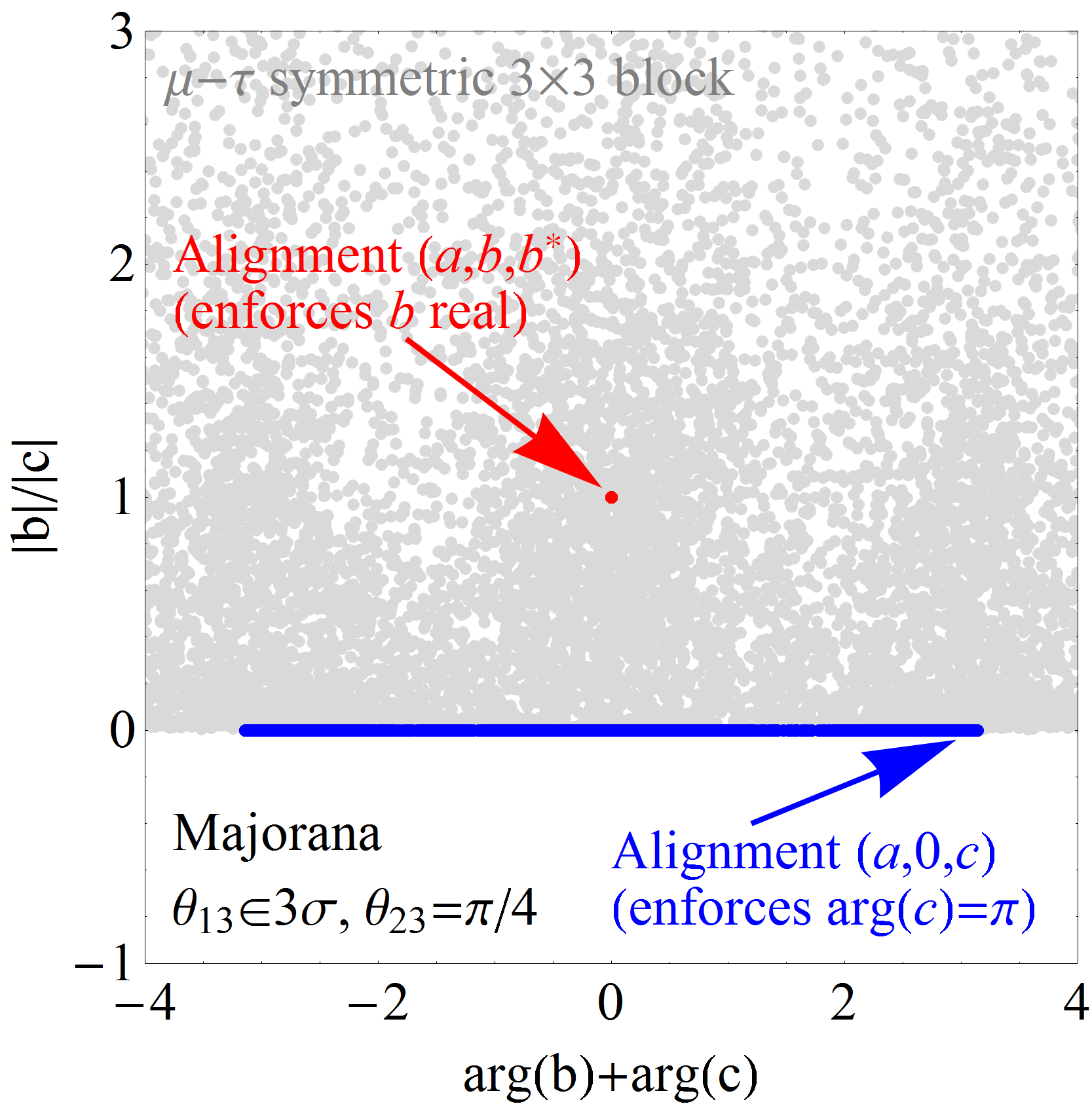} & \includegraphics[width=7cm]{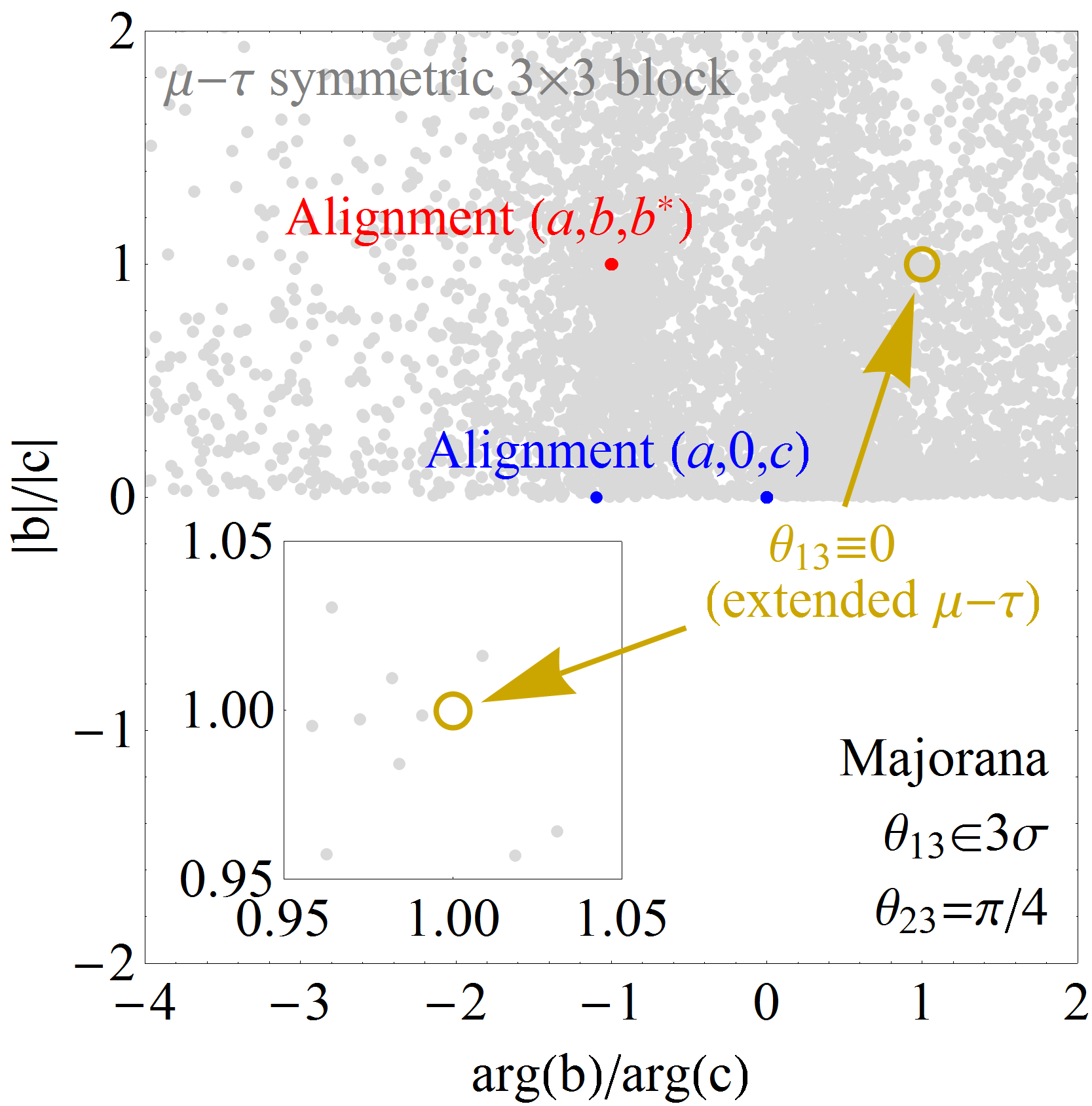}
\end{tabular}
\caption{\label{fig:ab_bc1}
Correlation between the complex parameters $b$ and $c$. In order to clearly reveal the different features, the same data are plotted in two different ways in the left and right columns, respectively; see main text for details. The Majorana phases are chosen to be zero in the upper row, and are varied in the lower row.
}
\end{figure}

A further interesting relation could potentially arise between the absolute values and the phases of the matrix elements $(a, b, c)$, which is displayed for $b$ and $c$ as examples in Fig.~\ref{fig:ab_bc1}. Let us again have a look at the upper panels first, for which the Majorana phases are all taken to be zero. Note that, in this figure, the same data set is displayed in two different ways in order to reveal certain features. Let us first look at the upper left panel. Here we plot the quantity ${\rm arg}(b) + {\rm arg}(c)$ versus the ratio $|b|/|c|$. As can be seen, the ball-park of the valid points requires that either ${\rm arg}(b) = - {\rm arg}(c)$ (i.e.\ if plotted in the complex plane and normalised to unit length, the two vectors $b$ and $c$ would transform into each other by a reflection on the real axis) or that $|b|=0$ (in which case the phase of $b$ is not well defined and thus ${\rm arg}(b) = - {\rm arg}(c)$ can be trivially fulfilled). This means in particular that these points \emph{cannot} be obtained by alignments such as $(a, b, c) = (1,4,2) a$~\cite{King:2013iva,King:2013xba}. There are also a few outlier points visible at phases $\pm \pi$. These values are in principle accessible (even though not ``likely'' from the parameter scan), as exemplified by the $D_4$-like (blue) alignment. Note that this alignment again does not enforce ${\rm arg}(c) = \pi$ by itself, but it does so when combined with $\mu - \tau$ symmetry. The $A_4$-like (red) alignment trivially imposes $|b|/|c|=1$, in which case ${\rm arg}(b) + {\rm arg}(c) = 0$ is enforced.

Now we turn to the upper right panel of Fig.~\ref{fig:ab_bc1}. Here, the same data set is displayed, however, this time as a function of ${\rm arg}(b)/{\rm arg}(c)$ instead of ${\rm arg}(b) + {\rm arg}(c)$. The reason is the following: as we had already mentioned, we need $b \neq c$ in order not to obtain a $4\times 4$ matrix with an extended $\mu - \tau$ symmetry, which would enforce $\theta_{13} \equiv 0$. Thus, if our numerical calculation is sensible, there should be no gray points found at $b = c$ or, equivalently, around the point $(|b|/|c|, {\rm arg}(b)/{\rm arg}(c)) = (1, 1)$. In the left panel, this point could not be displayed properly, since ${\rm arg}(b)={\rm arg}(c)$ would still allow for any value of ${\rm arg}(b) + {\rm arg}(c)$, but in the right panel it is marked in dark yellow. Indeed, although the same two branches of the correlations appear in the figure, no gray points are visible around $(1, 1)$, which is correct since none of them could possibly yield $\sin \theta_{13} = 0$. Note that, while this feature is clearly visible in that plot, the two alignments could not be displayed properly in the right panel: since the $A_4$-like (red) alignment together with the $\mu - \tau$ symmetry would force $b$ to be zero, the parameter ${\rm arg}(b)/{\rm arg}(b^*)$ would for those points essentially be a division of two (numerical) zeros in the case of varying Majorana phases. Similarly, for the $D_4$-like (blue) alignment, our numerical calculation would essentially find all kinds of values for ${\rm arg}(b)$, which would be meaningless since $|b|=0$, however, they would mess up the plot on the right panel. Indeed, for the information contained, there seems to be no optimum way to capture all the features in one single plot.

Unfortunately, nearly all these tendencies are again wiped out completely if the Majorana phases are taken to have general values, cf.\ lower panels of Fig.~\ref{fig:ab_bc1}. While some white patches may or may not be visible, there is certainly no correlation left for the gray points. For the red alignment, one can see that, in addition to $|b| = |c|$, we cannot only see that trivially ${\rm arg}(b) + {\rm arg}(c) = 0$ (lower left panel) or ${\rm arg}(b)/{\rm arg}(c) = -1$ (lower right panel), as we could have anticipated from $c = b^*$. However, the lower right panel reveals that, for the blue alignment, in addition to the trivial case ${\rm arg}(b) =0$, it could also be that ${\rm arg}(b)/{\rm arg}(c) = -1$. Unfortunately, this does not have any effect as long as $|b| = 0$. The only really solid prediction is that, even for the general case, the point $(1, 1)$ is still avoided by the gray dots. This is not easy to see by eye in the large version of the plot in the lower right panel, but the enlarged region in the inset shows that it is nevertheless correct.

\subsection{Alignments required to reproduce  $\nu_e$ disappearance results}

\begin{figure}[tp]
\centering
\begin{tabular}{lr}
\includegraphics[width=7.5cm]{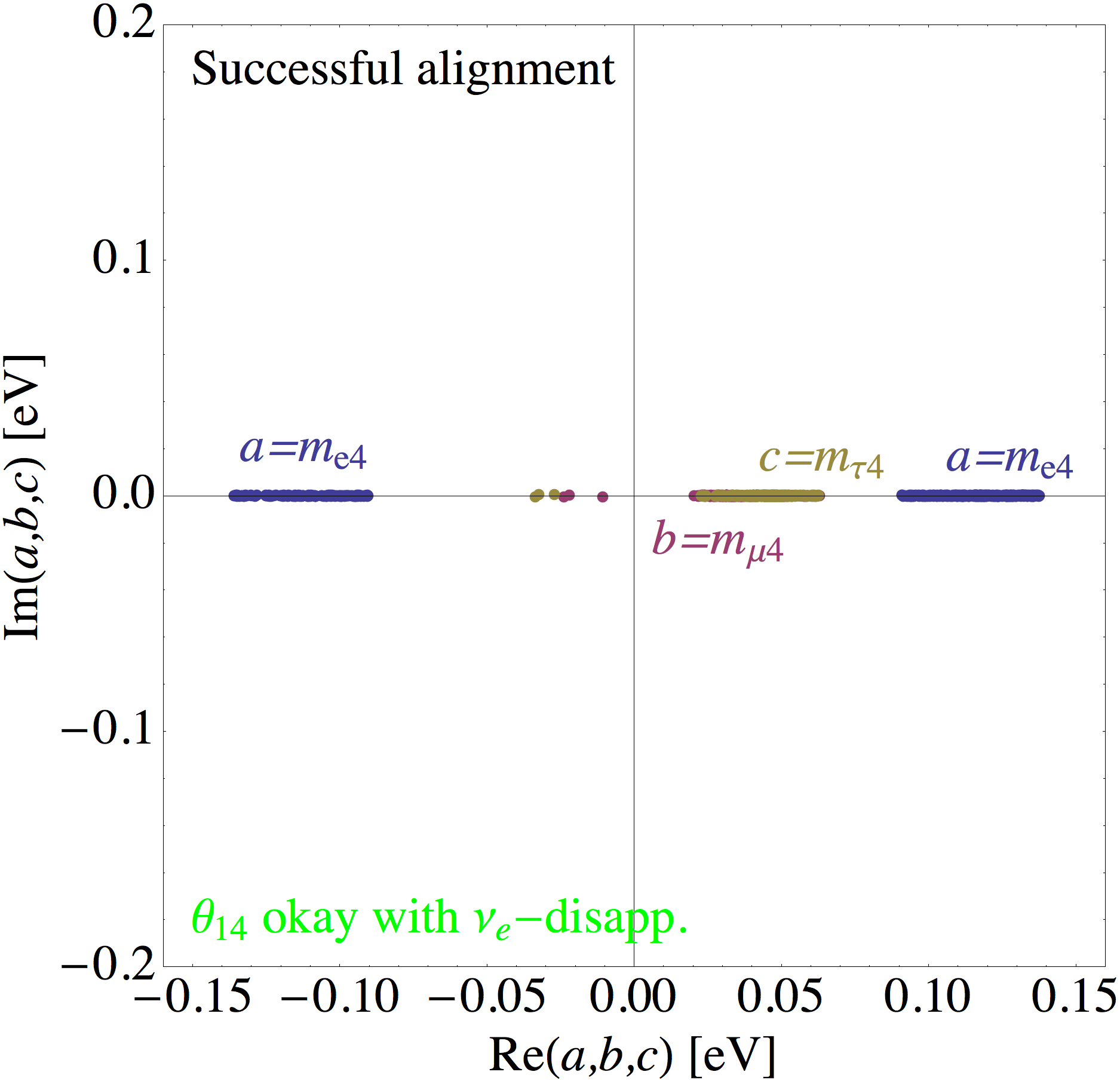} & \includegraphics[width=7cm]{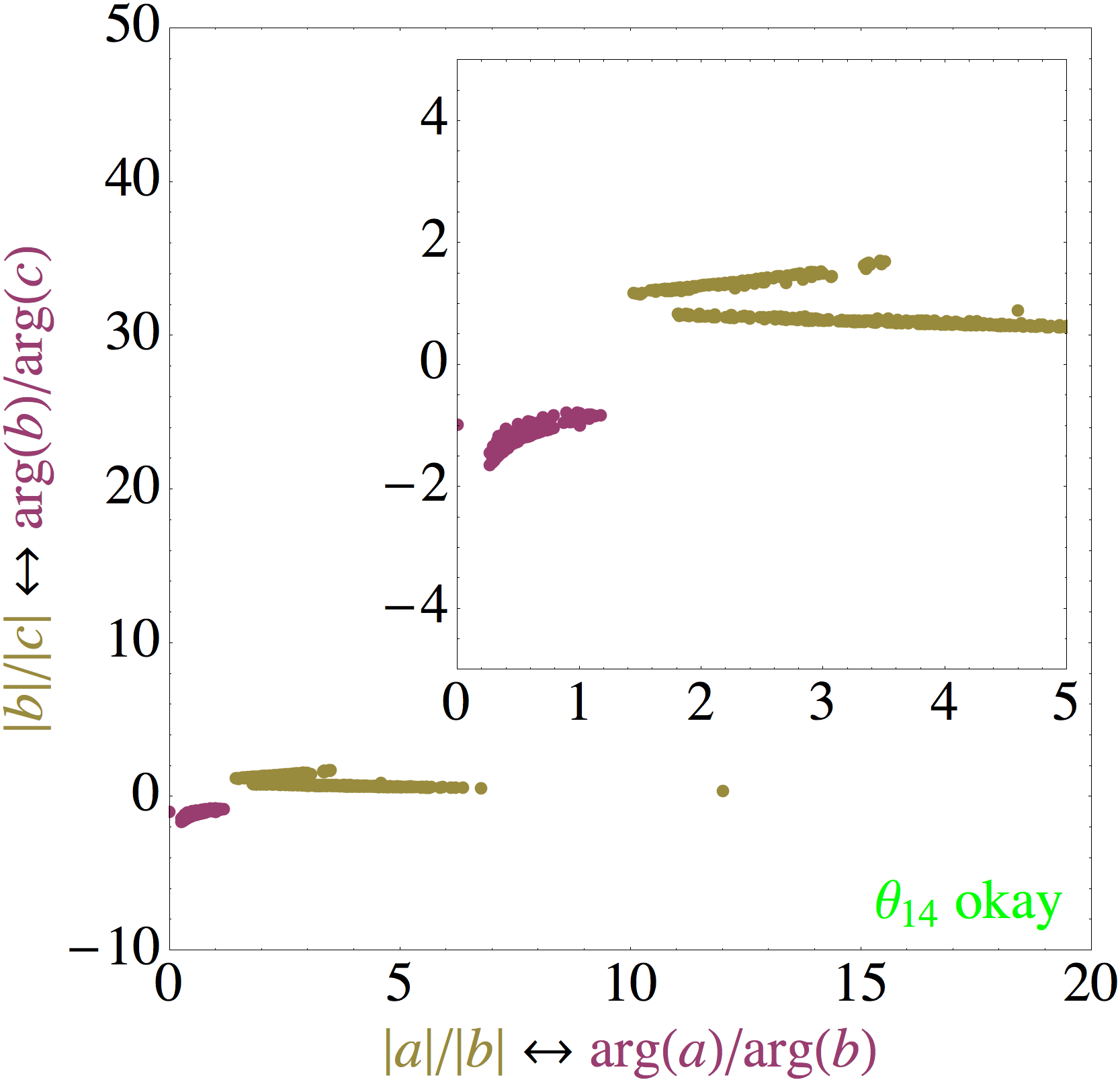}\\
\includegraphics[width=7.5cm]{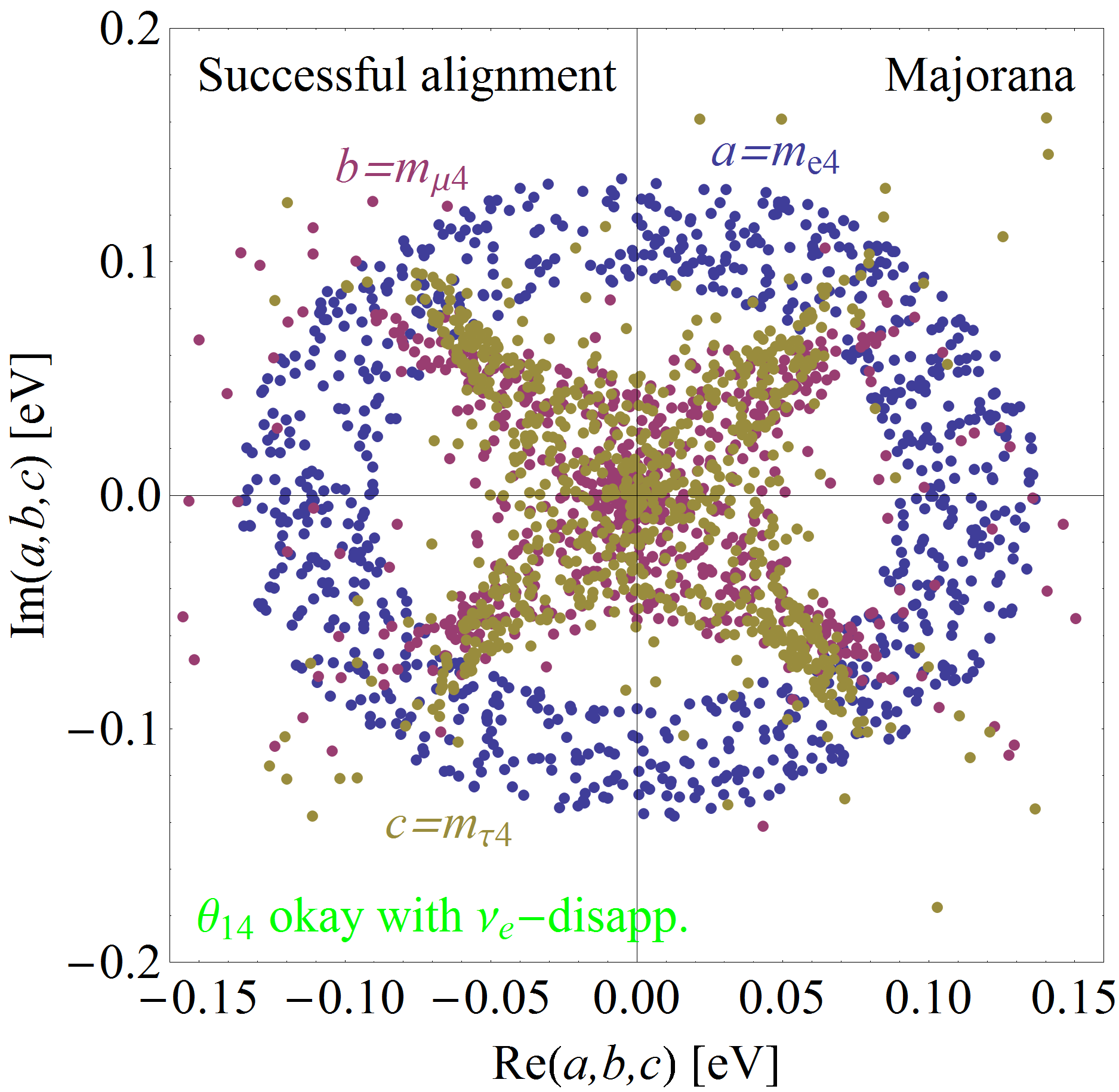} & \includegraphics[width=7cm]{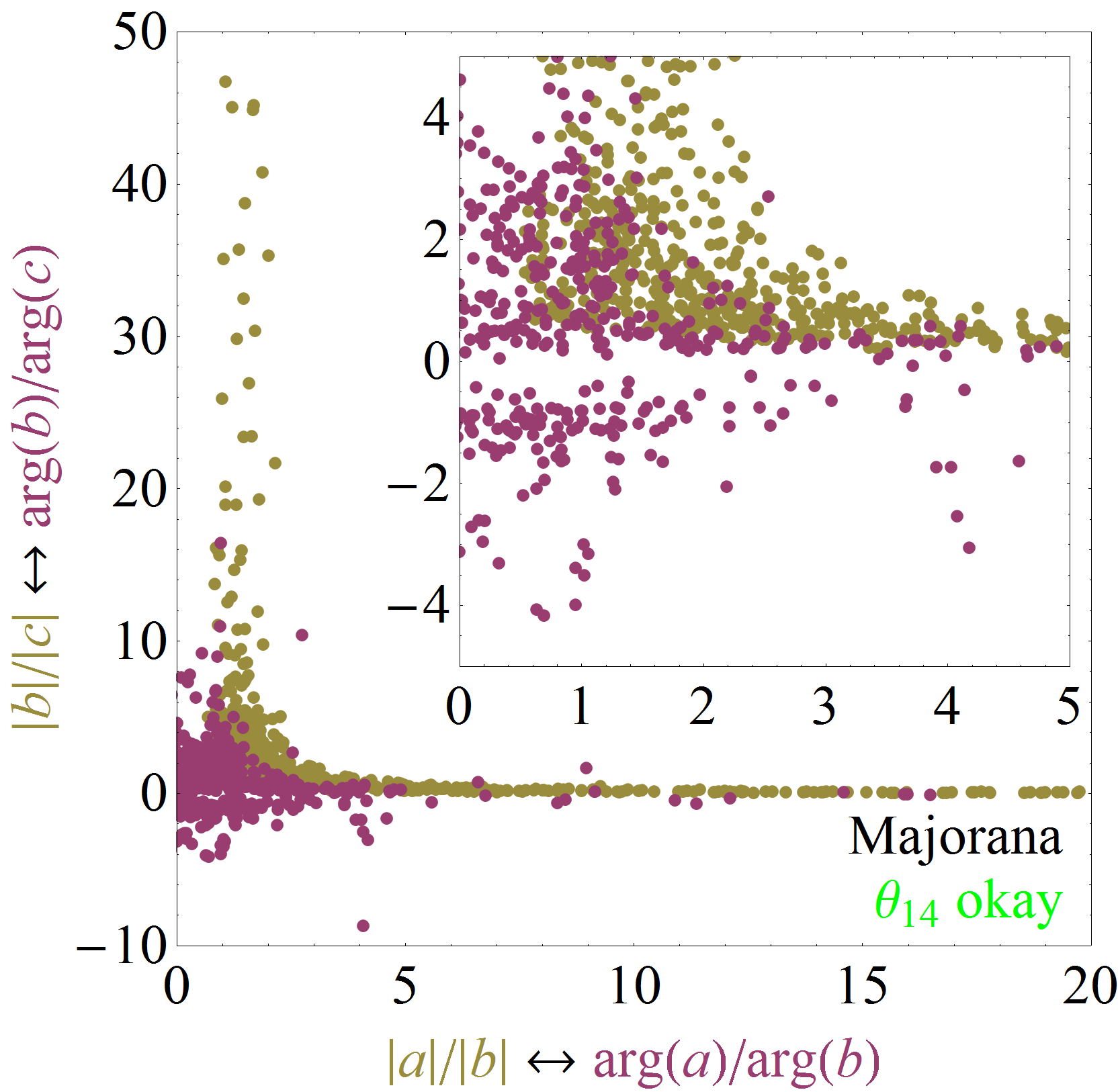}
\end{tabular}
\caption{\label{fig:abc_success}
Alignment points which can successfully reproduce the $\nu_e$-disappearance data. In the left column, we show the absolute sizes of the real and imaginary parts, where the different colour codings correspond to $a$, $b$, and $c$, respectively. In the right column, we show the ratios among different quantities, where the different colour codings correspond to the absolute values and phases, respectively. The Majorana phases are chosen to be zero in the upper row, and they are varied in the lower row. 
}
\end{figure}

Finally, we would like to ask the question which alignment $(a,b,c)$ we have to choose if we would like to successfully reproduce a certain part of the data. Since the null results only yield an upper bound and since our general (gray) region is incompatible with the combined appearance data as long as $\theta_{23}$ is taken to be maximal and the Majorana phases are taken to be zero, but it can easily fit the $\nu_e$ disappearance results, it would be interesting to see how $(a,b,c)$ have to be chosen for that data to be matched. This is shown in Fig.~\ref{fig:abc_success}, where we plot the absolute real and imaginary parts of $(a, b, c)$ on the left, and some ratios between moduli and arguments on the right. 

Starting with the upper left panel, it can be seen that for all points, the real parts dominate while the imaginary parts are comparatively small. Furthermore, while ${\rm Re}(a)$ can be positive or negative, ${\rm Re}(b,c)$ are practically always positive. Furthermore, there is a clear tendency for $|{\rm Re}(a)|$ to be considerably larger than ${\rm Re}(b,c)$. The latter two are practically of the same size, although a very slight tendency for ${\rm Re}(b) < {\rm Re}(c)$ is visible. Note that, since we display absolute elements of the neutrino mass matrix, all points given carry the unit eV. 

Allowing for the Majorana phases to vary reveals the actual correlation, cf\ lower left panel of Fig.~\ref{fig:13-generation1}. While the allowed regions for $b$ and $c$ form crosses that lie on top of each other (in fact, the corresponding points in the upper left panel also lie pratically on top of each other, which makes it a bit difficult to distinguish them visually), the points for a suitable entry $a$ lie on a circle around the origin with a radius of roughly $|a| \sim 0.1$~eV. Thus, while $b$ and/or $c$ can in principle be zero, the $e4$ element $a$ of the $4\times 4$ neutrino mass matrix must be non-zero with a well determined absolute value. Glancing at the upper left panel again, it is visible that setting the Majorana phases to zero essentially pics some of the regions of the circle and of the crosses which intersect (or, rather, are close to) the line with zero imaginary part, as to be expected from Eqs.~\eqref{eq:bc}. However, in addition the points where $b,c > 0$ are much more likely in that case, which is a feature that is non-trivial to understand.

On the upper right panel, in turn, we instead show certain ratios of quantities, namely $|b|/|c|$ vs.\ $|a|/|b|$ (dark yellow points) and ${\rm arg}(b)/{\rm arg}(c)$ vs.\ ${\rm arg}(a)/{\rm arg}(b)$ (purple points), again for vanishing $(\alpha, \beta, \gamma)$. Also here, clear correlations are visible. In particular there is a tendency for the phases of $b$ and $c$ to have different signs, while the phases of $a$ and $b$ always have the same sign. Furthermore, the inset shows a region where ${\rm arg}(a)/{\rm arg}(b) \ll 1$ while ${\rm arg}(b)/{\rm arg}(c) \simeq -0.6$, which simply means that for these points both ${\rm arg}(b,c)$ are very small (i.e., $b$ and $c$ are nearly real), but there is a fixed ratio between the two arguments.

Allowing the Majorana phases to vary, cf.\ lower right panel, again increases the allowed regions considerably. However, at least some general tendencies are visible, namely that $|b|$ should be somewhat small, unless $|a|$ is small, and quite generally most of the poitns cluster around $(a, b, c)$ having non-identical values which are however of the same order of magnitude.\\

Further such tendencies could be read off this plot and, hopefully, they will give an indication to model builders where in the parameter space to look for a prediction that yields an active-sterile mixing in the correct region.

\section{\label{sec:models}Ideas for model building}

In the literature there are many interesting examples of models giving a $\mu-\tau$ symmetric neutrino mass matrix in the basis where the charged leptons are diagonal. Here we consider two examples to show how  we can apply our results. One of the first example is based on $A_4$~\cite{Ma:2001dn,Babu:2002dz} and the second one on a $D_4$~\cite{Grimus:2003kq} flavour symmetry. We shortly describe the main features of both models and we show how we can extend them by a sterile neutrino in order to generate the reactor angle.

\subsection{$A_4$ model}

The model from Ref.~\cite{Babu:2002dz} is supersymmetric and it is based on the flavour symmetry $A_4$. This is the finite group of even permutations of four objects. It has three singlet and one triplet irreducible representations, and it is the smallest non-Abelian discrete group featuring triplets.

\begin{table}[t!]
\begin{center}
\begin{tabular}{|c|c|c|c|c|c|c|c|c|c|c|c|c|c|}
\hline
&$\hat Q$ & $\hat L$ & $\hat u^c_1, ~\hat d^c_1, ~\hat e^c_1$ &$ \hat u^c_2, ~\hat d^c_2, ~\hat e^c_2 $&$ \hat u^c_3, ~\hat d^c_3, ~\hat e^c_3$ & $\hat \phi_{1,2}$ &$\hat U$&$\hat U^c$ &$\hat D$ &$\hat D^c$&$\hat E$&$\hat E^c$  &$\hat \chi$\\
\hline
$A_4$ & $\mathbf{3}$ & $\mathbf{3}$ & $\mathbf{1}$ &$\mathbf{1'}$ & $\mathbf{1''}$& $\mathbf{1}$ & $\mathbf{3}$ & $\mathbf{3}$ & $\mathbf{3}$ &$\mathbf{3}$ & $\mathbf{3}$& $\mathbf{3}$ & $\mathbf{3}$\\
\hline
$Z_3$ & $1$ & $1$ & $\omega$ &$\omega$ & $\omega$& $1$ & $1$ & $1$ & $1$& $1$&$1$& $1$& $\omega^2$\\
\hline
\end{tabular}
\caption{\label{tabbmv}Matter assignment of the model of Ref.\,\cite{Babu:2002dz}. Note that $\omega^3 = 1$ and $1 + \omega + \omega^2 = 0$.}
\end{center}
\end{table}

The usual quark, lepton, and Higgs superfields transform under $A_4$ as detailed in Tab.~\ref{tabbmv}, where extra heavy $SU(2)$ singlet quark, lepton, and Higgs superfields are also added. The superpotential  is  given by
\begin{eqnarray}
\hat W &=& M_U \hat U_i \hat U^c_i + f_u \hat Q_i \hat U^c_i \hat \phi_2 + h^u_{ijk} \hat U_i \hat u^c_j \hat \chi_k + M_D \hat D_i \hat D^c_i + f_d \hat Q_i \hat D^c_i \hat \phi_1 + h^d_{ijk} \hat D_i \hat d^c_j \hat \chi_k \nonumber \\
&& + M_E \hat E_i \hat E^c_i + f_e \hat L_i \hat E^c_i \hat \phi_1 + h^e_{ijk} \hat E_i \hat e^c_j \hat \chi_k + \mu \hat \phi_1 \hat \phi_2 \nonumber \\
&& + \frac{1}{2} M_\chi \hat \chi_i \hat \chi_i + h_\chi \hat \chi_1 \hat \chi_2 \hat \chi_3.
\end{eqnarray}
The $Z_3$ auxiliary symmetry is explicitly broken softly by $M_\chi \neq 0$. The scalar potential for the fields $\chi_i$ is given by
\begin{equation}
V = |M_\chi \chi_1 + h_\chi \chi_2 \chi_3|^2 + |M_\chi \chi_2 + h_\chi \chi_3
\chi_1|^2 + |M_\chi \chi_3 + h_\chi \chi_1 \chi_2|^2,
\end{equation}
and from its minimisation we get:
\begin{equation}
\langle \chi_1 \rangle = \langle \chi_2 \rangle = \langle \chi_3 \rangle =
u = -M_\chi/h_\chi.
\end{equation}
Consider now the $6 \times 6$ Dirac mass matrix linking $(e_i,E_i)$ to
$(e_j^c,E_j^c)$,
\begin{equation}
{\cal M}_{eE} = \left( \begin{array} {c@{\quad}c@{\quad}c@{\quad}c@{\quad}c@
{\quad}c} 0 & 0 & 0 & f_e v_1 & 0 & 0 \\ 0 & 0 & 0 & 0 & f_e v_1 & 0 \\
0 & 0 & 0 & 0 & 0 & f_e v_1 \\ h_1^e u & h_2^e u & h_3^e u & M_E & 0 & 0 \\
h_1^e u & h_2^e \omega u & h_3^e \omega^2 u & 0 & M_E & 0 \\
h_1^e u & h_2^e \omega^2 u & h_3^e \omega u & 0 & 0 & M_E \end{array} \right),
\end{equation}
where $v_1 = \langle \phi_1^0 \rangle$. The quark mass matrices look similar. The reduced $3 \times 3$ charged lepton mass matrix is 
\begin{equation}
{\cal M}_e = U_L \left( \begin{array} {c@{\quad}c@{\quad}c} {h_1^e}' & 0 & 0
\\ 0 & {h_2^e}' & 0 \\ 0 & 0 & {h_3^e}' \end{array} \right) \frac{\sqrt{3} f_e v_1 u}{M_E},
\end{equation}
where ${h_i^e}' \equiv h_i^e [1+(h_i^e u)^2/M_E^2]^{-1/2}$ and
\begin{equation}
U_L = \frac{1}{\sqrt{3}} \left( \begin{array} {c@{\quad}c@{\quad}c} 1 & 1 & 1
\\ 1 & \omega & \omega^2 \\ 1 & \omega^2 & \omega \end{array} \right).
\end{equation}

Clearly, the \emph{up} and \emph{down} quark mass matrices are obtained in the same way and are both diagonalised by $U_L$, so that the charged-current mixing Cabibbo-Kobayashi-Maskawa (CKM) matrix $V_{\rm CKM}$ is the identity matrix. The small measured CKM angles may be generated from corrections associated to the structure of the soft supersymmetry breaking sector to make the model viable.

In this model the neutrino masses arise from the dimension-5 Weinberg operator,
\begin{equation}
\frac{f_\nu}{\Lambda} \hat L_i \hat \phi_2 \hat L_i \hat \phi_2.
\end{equation}

The effective Majorana neutrino mass matrix in the basis where the charged lepton mass matrix is diagonal is given by
\begin{equation}\label{mnuA4p}
{\cal M}_\nu = \frac{f_\nu v_2^2}{\Lambda} U_L^T U_L = \frac{f_\nu^2 v_2^2}{\Lambda}
\left( \begin{array} {c@{\quad}c@{\quad}c} 1 & 0 & 0 \\ 0 & 0 & 1 \\
0 & 1 & 0 \end{array} \right)
\equiv \frac{f_\nu v_2^2}{\Lambda} \lambda^0,
\end{equation}
giving (at this stage) a maximal atmospheric mixing angle but degenerate light neutrino masses.
 
Going down to the electroweak scale, $\lambda^0$ in Eq.~\eqref{mnuA4p} is corrected by the wave-function renormalisations of $\nu_e$, $\nu_\mu$, and $\nu_\tau$, as well as by the corresponding vertex renormalisations, i.e.\ $\lambda^0\to \lambda$, which breaks the neutrino mass degeneracy. The radiative corrections associated with a general slepton mass matrix in softly broken supersymmetry (related to $\nu_i \to \nu_j$ transitions) are given by the matrix
\begin{equation}
R = \left( \begin{array} {c@{\quad}c@{\quad}c} 
\delta_{ee} & \delta_{e \mu}  &  \delta_{e \tau} \\
\delta_{e \mu}  & \delta_{\mu \mu} &  \delta_{\mu \tau} \\ 
\delta_{e \tau} & \delta_{\mu \tau} & \delta_{\tau\tau} \end{array} \right),
\end{equation}
so that at the low scale $\lambda$ is:
\begin{equation}
\lambda = \lambda^0+ R \lambda^0+\lambda^0 R^T= 
\left( \begin{array} {c@{\quad}c@{\quad}c} 1 + 2 \delta_{ee} &
\delta_{e \mu} + \delta_{e \tau} & \delta_{e \mu} + \delta_{e \tau} \\
\delta_{e \mu} + \delta_{e \tau} & 2 \delta_{\mu \tau} & 1 + \delta_{\mu \mu}
+ \delta_{\tau \tau} \\ \delta_{e \mu} + \delta_{e \tau} & 1 + \delta_{\mu \mu}
+ \delta_{\tau \tau} & 2 \delta_{\mu \tau} \end{array} \right),
\end{equation}
where we have assumed all parameters to be real for simplicity. The above mass matrix, ${\cal M}_\nu$, is clearly $\mu-\tau$ invariant and yields a zero reactor angle as well as maximal atmospheric mixing.

In order to generate  a non-zero reactor angle in this model, we can use the method described in this paper that makes use of one sterile neutrino $\hat \nu_s$ which transforms as a singlet under $A_4$. We assume that the sterile neutrino is charged under an extra auxiliary symmetry $Z_2$. We also add to the particle content of the model a scalar electroweak singlet flavon $\xi$ that is charged under $Z_2$(this parity ensures that the flavon $\xi$ can glue only to the sterile neutrino) and that transforms as a triplet under $A_4$: $\xi = (\xi_1, \xi_2, \xi_3)$. Thus, the superpotential contains the following extra term that mixes the active and sterile neutrinos and also adds a sterile neutrino mass term,
\begin{equation}
\hat W \supset \frac{f_s}{\Lambda} \hat L_i \hat \phi_2 \nu_s \xi_i+ \frac{m_s}{2} \hat \nu_s\hat \nu_s,
\end{equation}
where $\Lambda$ is an effective scale.

The fourth column of the neutrino mass matrix is then proportional to the VEVs of the flavons $\xi_i$, giving
\begin{equation}
a=\frac{f_s}{\Lambda} v_2 \langle \xi_1 \rangle\,,\quad b=\frac{f_s}{\Lambda} v_2 \langle \xi_2 \rangle\,,\quad
c=\frac{f_s}{\Lambda} v_2 \langle \xi_3 \rangle\,.\quad
\end{equation}

From the model-independent numerical analysis in Secs.~\ref{sec:method} to~\ref{sec:theory} it is clear that the two scalar $A_4$ triplets $\xi$ and $\chi$ must take VEVs in different directions of $A_4$,
\begin{equation}
\langle \chi \rangle\sim (1,1,1)\ne (a,b,c) \sim \langle \xi \rangle .
\end{equation}
It is well-known that, given two different $A_4$ scalar triplets $\xi$ and $\chi$, the minimisation of their scalar potential $V(\xi,\chi)$ yields as a natural solution  $\langle \chi \rangle \sim \langle \xi \rangle$, i.e., the VEVs of the two fields are aligned. This is in contrast to the requirement obtained from our numerical results, because we had found that  the two triplets must take VEVs in different $A_4$-directions. Typically, in order to solve such a problem, it is required to break the flavour symmetry explicitly in the scalar potential or to make use of extra dimensions. It is not the purpose of this paper to give a complete model, but just to suggest possible strategies that could be followed. Using  explicit $A_4$ breaking terms it is quite straightforward to obtain the VEV misalignment required, and we do not embark this enterprise in more detail. We also want to comment on the possibility to use extra dimensions. In this case we could assume that, following the general idea of~\cite{Altarelli:2005yp}, $\nu_s$ and $\xi$ live on the $y=L$ ultraviolet (UV) brane while all the other fields stay on the Standard Model (SM) $y=0$ brane. Since in this framework $\chi$ and $\xi$ are located on different branes, their potentials are separated and can easily have independent minima. However the sets of scalar fields that live on different branes can interact at higher order, giving a deviation of the vacuum alignments. But such a deviation is typically of order $1/(\Lambda\,L)^4$ (see \cite{Altarelli:2005yp} for a detailed discussion), where $\Lambda$ is the effective scale of the model. It is clear that, for sufficiently large $L$, the vacuum alignment corrections are negligible. A detailed study of these corrections is beyond the scope of the present paper because we have only sketched some possible ideas, while for a complete study it is necessary to fix a particular model.

\subsection{$D_4$ model}

This model  is based on the dihedral group $D_4$~\cite{Grimus:2003kq}\footnote{The dihedral group $D_4$ is isomorphic to the group of permutation of three objects $S_3$.} which has five irreducible representations, four singlets $\mathbf{1}_{++}$, $\mathbf{1}_{+-}$, $\mathbf{1}_{-+}$, $\mathbf{1}_{--}$, and one doublet $\mathbf{2}$. The product of two doublets is $\mathbf{2} \otimes \mathbf{2} = \mathbf{1}_{++} \oplus \mathbf{1}_{+-} \oplus \mathbf{1}_{-+} \oplus \mathbf{1}_{--}$, and the products of the singlets are trivial (for example, $\mathbf{1}_{+-} \otimes \mathbf{1}_{-+} = \mathbf{1}_{--}$). Differently from the previous one, this model is not supersymmetric. The SM is only extended by adding three right-handed neutrinos $\nu^c_{1,2,3}$, three Higgs doublets $H_{1,2,3}$, and two neutral scalar singlets $\chi_{1,2}$, as detailed in Tab.~\ref{tab:D4}.

\begin{table}[t!]
\begin{center}
\begin{tabular}{|c|c|c|c|c|c|c|c|c|c|c|}
\hline
 & $L_e$ & $e^c$ & $L_{\mu,\tau} $ & $\mu^c,\tau^c$ & $\nu^c_1$ & $\nu^c_{2,3}$ & $H_1$ & $H_{2}$ & $H_{3}$ & $\chi_{1,2}$  \\
\hline
$D_4$ & $\mathbf{1}_{++}$ & $\mathbf{1}_{++}$ & $\mathbf{2}$ & $\mathbf{2}$ & $\mathbf{1}_{++}$ & $\mathbf{2}$ & $\mathbf{1}_{++}$ & $\mathbf{1}_{++}$ & $\mathbf{1}_{+-}$ & $\mathbf{2}$\\
$Z_2^{\rm aux}$ & $+$ & $-$ & $+$ & $+$ & $-$ & $-$ & $-$ & $+$ & $-$ & $+$ \\
\hline
\end{tabular}
\caption{\label{tab:D4}Matter content of the model from Ref.~\cite{Grimus:2003kq}.}
\end{center}
\end{table}
The Lagrangian is given by 
\begin{eqnarray}
\mathcal{L}&=&[y_1 \overline{L_e} \nu^c_1 +y_2(\overline{L_\mu} \nu^c_2+ \overline{L_\tau} \nu^c_3)] \tilde{H}_1+\nonumber \\
&& +y_3  \overline{L_e} e^c_1 H_1 +y_4 (\overline{L_\mu} \mu^c+ \overline{L_\tau} \tau^c) H_2 +y_5( \overline{L_\mu} \mu^c- \overline{L_\tau} \tau^c) H_3+\nonumber \\
&& + y_\chi{\nu_1^c}^T (\nu_2^c \chi_1+\nu_2^c \chi_2)+ M {\nu_1^c}^T \nu_1^c  + 
M' ({\nu_2^c}^T \nu_2^c +{\nu_3^c}^T \nu_3^c ) + H.c.
\end{eqnarray}
After flavour symmetry breaking, the $\chi$-fields take VEVs $\langle\chi_1\rangle =\langle\chi_2\rangle $, giving a $\mu - \tau$ invariant neutrino mass matrix (with maximal atmospheric and zero reactor mixings), while the charged lepton mass matrix is diagonal. Here we do not give other details and we refer interested readers to the original paper. 

Like in the previous case based on the $A_4$ group, we can generate a deviation of the reactor angle from zero by the use of a sterile neutrino $\nu_s$. It is also required to introduce three extra scalar fields, $\xi_{1}\sim \mathbf{1}_{++}$ and $(\xi_2,\xi_3)\sim \mathbf{2}$ under $D_4$, which are gauge singlets. Then, the following new terms are allowed in the Lagrangian:
\begin{equation}
\mathcal{L} \supset 
\frac{f_{s1}}{\Lambda}  \overline{L_e} \tilde{H}_1 \nu_s \xi_1+ \frac{f_{s2}}{\Lambda} [  \overline{L_\mu}   \xi_2  + \overline{L_\tau}   \xi_3] \tilde{H}_1 \nu_s +
\frac{m_s}{2}  \overline{\nu_s^c} \nu_s + H.c.
\end{equation}
As in the $A_4$ case, we assume that the fields $\nu_s$ and $\xi_{1,2,3}$ transform non-trivially under an extra $Z_2$ symmetry. In order to generate the reactor angle, the VEV of the $D_4$-doublet $(\xi_2,\xi_3)$ must break the $\mu-\tau$ symmetry, by $\langle \xi_2\rangle \ne \langle \xi_3\rangle$. This may be in contrast to the alignment $\langle\chi_1\rangle =\langle\chi_2\rangle $. Such misalignment problems can again be solved easily by using extra dimensions, just like in the $A_4$ case. A detailed study of this possibility goes beyond the scope of the present paper.

\section{\label{sec:conc}Summary and conclusions}

In this paper we have considered the possibility that the recently measured reactor angle and the active-sterile mixings, needed to describe the short-baseline anomalies, have a common origin. This is suggested from the fact that the active-sterile mixings obtained in fits of the short-baseline data in $3+N$ models are of the same order as the reactor angle. We have assumed the simplest framework possible, with only one sterile neutrino (giving a $4\times 4$ neutrino mass matrix). We have postulated that the reactor neutrino mixing vanishes in the active-active mass matrix part, which is why we have considered the $3\times 3$ active neutrino mass matrix to be $\mu-\tau$ invariant. This assumption implies  that the atmospheric mixing angle is almost maximal, in compatibility with data. As a consequence, both a non-zero value of $\theta_{13}$ and the active-sterile mixings originate from the active-sterile mass matrix elements and are potentially of the same order of magnitude. 

There have been several important questions of our analysis: 1) Which correlations among or constraints on the observables are implied in this framework?, 2) Can the short-baseline anomalies be reproduced?, 3) What are the requirements for the vacuum alignments of the VEVs?, and 4) What does that imply for flavour models?

We have demonstrated that $\theta_{14}$, which in our parameterisation leads to electron neutrino disappearance, must be non-zero in this framework. On the other hand, either $\theta_{24}$, leading to muon neutrino disappearance, or $\theta_{34}$ can vanish (but not both at the same time -- they are anti-correlated). Therefore, this framework is perfectly consistent with the reactor and gallium anomalies, and with the non-observation of muon neutrino disappearance. It is more difficult to reconcile this approach with the LSND results, as this is possible only for specific choices of the Majorana phases.

We have also shown how the active-sterile mixing and the non-zero value of $\theta_{13}$ emerge from the misalignment of the active-sterile VEVs, i.e., the explicit breaking of the $\mu - \tau$ symmetry. We have noted that ``misalignment'' could refer to the absolute values and/or phases of the VEVs, which can both be the origin of the breaking of the $\mu - \tau$ symmetry. We have also demonstrated that specific assumptions for the alignments, which can be found in the literature based on $A_4$ and $D_4$ models, are in fact very predictive. These choices may also impact the predictions for neutrinoless double beta decay. A detailed study is beyond the scope of this work, as the phenomenology of neutrinoless double beta can considerably change in the presence of light sterile neutrinos, see e.g.\ Refs.~\cite{Barry:2011wb,Girardi:2013zra,Merle:2013ibc}.

As far as the implications for flavour models are concerned, we have sketched the requirements in terms of two well-known example models based on $A_4$ and $D_4$, respectively. For instance, in the $A_4$ model, scalar electroweak singlet flavons are needed which must be triplets under $A_4$ to generate neutrino masses. It is however well-known that it is difficult for these triplets to take VEVs in different directions of $A_4$. We have proposed either an explicit breaking of $A_4$ or the use of extra spatial dimensions. In the latter case, the sterile neutrino and one of the flavon triplets would live on the UV brane, whereas the other flavon and SM fields reside on the infrared/SM brane. 

We conclude that, if sterile neutrinos exist, it is possible for active-sterile mixings and the non-zero value of $\theta_{13}$ to have a common origin in terms of flavour models. While we have studied the simplest setting possible, models with more than one sterile neutrino may have much wider possibilities. Our starting point has been the $\mu - \tau$ symmetric case, but other possibilities are viable as well -- as for example tri-bimaximal mixing. In such alternative approaches, it may also be possible to describe a non-zero $\theta_{13}$ and strong deviations from maximal atmospheric mixing at the same time, whereas our framework has implied $\theta_{23}$ being close to maximal.

\section*{Acknowledgments}

A.M.\ acknowledges support by a Marie Curie Intra-European Fellowship within the 7th European Community Framework Programme FP7-PEOPLE-2011-IEF, contract PIEF-GA-2011-297557, and partial support from the European Union FP7 ITN-INVISIBLES (Marie Curie Actions, PITN-GA-2011-289442). S.M.\ and W.W.\ thank the DFG grants WI 2639/3-1 and WI 2639/4-1 for financial support.

\bibliographystyle{apsrev}
\bibliography{references}

\end{document}